\documentclass[twocolumn]{autart} 
\usepackage{amsmath,bm}
\usepackage{amsfonts}
\usepackage{amssymb}
\usepackage[final]{graphicx}
\usepackage{mathtools}
\usepackage{upgreek,booktabs}
\usepackage{mathrsfs}
\usepackage{algorithm,algpseudocode}
\usepackage{paralist}
\usepackage[hidelinks,colorlinks=false]{hyperref}
\usepackage{tikz,pgfplots}
\usepackage{footnote}
\makesavenoteenv{tabular}
\makesavenoteenv{table}
\usetikzlibrary{arrows}
\usetikzlibrary{positioning}
\usepackage{subcaption}
\newtheorem{theorem}{Theorem}

\newtheorem{proposition}{Proposition}

\newtheorem{definition}{Definition} 
\newtheorem{remark}{Remark}
\usepackage{amsmath}
\usepackage{amssymb}
\newcommand{\Tpiy}{T(\pi,y)}
\newcommand{\Spiy}{\sigma(\pi,y)}

\newcommand{\Root}{X}
\newcommand{\Obs}{Y}
\renewcommand{\prob}{\mathbb{P}}
\newcommand{\state}{X}
\newcommand{\Time}{t}
\newcommand{\statedim}{S}

\newcommand{\statespace}{\mathcal{S}}
\newcommand{\tp}{P}
\newcommand{\belief}{\pi}
\newcommand{\Belief}{\Pi}
\newcommand{\action}{u}
\newcommand{\obspace}{\mathcal{Y}}
\newcommand{\obp}{B}
\newcommand{\filter}{T}
\newcommand{\history}{Z}
\newcommand{\E}{\mathbb{E}}
\newcommand{\obs}{Y}
\newcommand{\obsy}{y}
\newcommand{\filternorm}{\sigma}
\newcommand{\one}{\textbf{1}}

\newcommand{\policy}{\mu}
\newcommand{\policyspace}{\mathcal{U}}
\newcommand{\actionspace}{\mathcal{A}}
\newcommand{\discount}{\rho}
\newcommand{\discountedreward}{J}
\DeclareMathOperator*{\argmax}{\arg\,\max\;}
\newcommand{\optpolicy}{\policy^*}

\newcommand{\valuefunction}{V}

\newcommand{\looks}{l}

\newcommand{\reals}{{\rm I\hspace{-.07cm}R}}
\newcommand{\reward}[1] {r^\prime #1}
\newcounter{daggerfootnote}

\newcounter{ddaggerfootnote}

\newcommand{\expnumber}[2]{{#1}\mathrm{e}{#2}}
\let\oldReturn\Return
\renewcommand{\Return}{\State\oldReturn}
\usepackage{enumitem}
\makeatletter
\newcommand{\mylabel}[2]{#2\def\@currentlabel{#2}\label{#1}}
\makeatother
\makeatletter
\def\munderbar#1{\underline{\sbox\tw@{$#1$}\dp\tw@\z@\box\tw@}}
\makeatother
\newcommand\given[1][]{\:#1\vert\:}
\usepackage[export]{adjustbox}
\maketitle
\begin{document}
\begin{frontmatter}

\title{Multiple Stopping Time POMDPs: Structural Results \& Application in Interactive Advertising in Social Media}
\author{Vikram Krishnamurthy$^{\dagger}$, Anup Aprem$^{\ddagger}$ and Sujay Bhatt$^{\dagger}$
}
\thanks{$^{\dagger}$ V.~Krishnamurthy and S.~Bhatt are with the Department of Electrical \& Computer Engineering and Cornell Tech, Cornell University, NY }%
\thanks{$^{\ddagger}$ A.~Aprem is with the Department of Electrical \& Computer Engineering, University of British Columbia, Vancouver, BC, Canada }
\thanks{E-mail addresses: vikramk@cornell.edu (V.~Krishnamurthy), aaprem@ece.ubc.ca (A.~Aprem), sh2376@cornell.edu (S.~Bhatt)}
\thanks{A significantly shortened version of this paper containing partial results appeared in: V.~Krishnamurthy, A.~Aprem and S.~Bhatt, ``Multiple stopping time POMDPs: Structural results,'' 2016 54th Annual Allerton Conference on Communication, Control, and Computing (Allerton), Monticello, IL, 2016, pp. 115-120.}

\begin{abstract}
This paper considers a multiple stopping time problem for a Markov chain observed in noise, where a decision maker chooses at most $L$ stopping times to maximize a cumulative objective. We formulate the problem as a Partially Observed Markov Decision Process (POMDP) and derive structural results for the optimal multiple stopping policy. The main results are as follows: 
\begin{inparaenum}[i)] 
\item The optimal multiple stopping policy is shown to be characterized by threshold curves $\Gamma_l$, for $l = 1,\cdots, L$, in the unit simplex of Bayesian Posteriors. 
\item The stopping sets $S^{l}$ (defined by the threshold curves $\Gamma_l$) are shown to exhibit the following nested structure $\mathcal{S}^{l-1} \subset \mathcal{S}^{l}$. 
\item The optimal cumulative reward is shown to be monotone with respect to the copositive ordering of the transition matrix. 
\item A stochastic gradient algorithm is provided for estimating linear threshold policies by exploiting the structural results. These linear threshold policies approximate the threshold curves $\Gamma_{l}$, and share the monotone structure of the optimal multiple stopping policy.     
\end{inparaenum}
As an illustrative example, we apply the multiple stopping framework to interactively schedule advertisements in live online social media. It is shown that advertisement scheduling using multiple stopping performs significantly better than currently used methods. 
\end{abstract}

\end{frontmatter}
\section{Introduction}
Classical optimal stopping time problem is concerned with choosing a single time to take a stop action by observing a sequence of random variables in order to maximize a reward function. 
It has applications in numerous fields ranging from hypothesis testing~\cite{Lai97,Lai01}, parameter estimation~\cite{Lai01}, machine replacement~\cite{Rus87,Mon80}, multi-armed bandits and quickest change detection~\cite{poor2009quickest,Kri11,KB16}. 
The optimal multiple stopping time problem generalizes the classical single stopping problem; the objective is to stop $L$-times to maximize the cumulative reward. 

\begin{figure}[htb]
\centering
\tikzstyle{int}=[]
\tikzstyle{init} = [pin edge={to-,thin,black}]
\begin{tikzpicture}[auto,>=latex']
	\node [draw, text width=10em, label={Live Session}, align=center] (a) {Broadcaster {\small(Stochastic~Scheduler)}};
    \node (b) [below of=a,node distance=2.5em, coordinate] {};
    \node (c) [right of=b,node distance=5em, coordinate] {};
    \node (d) [left of=b,node  distance=5em, coordinate] {};
    \draw[-] (a) to (b);
    \draw[-] (b) to (d);
    \draw[-] (b) to (c);
    \node [draw, below of=d, minimum width=4em] (f) {Live Video};
    \node [draw, below of=c, minimum width=4em] (e) {Schedule Ads};
    \draw[->] (d) --  node[anchor=center, left, midway] {Continue} (f);
    \draw[->] (c) --  node[anchor=center, right, midway] {Stop} (e);
    \node (g) [below of=e,node distance=4em, coordinate] {};
    \node (h) [below of=f,node distance=4em, coordinate] {};
    \draw (h) -- coordinate[midway](m) (g) ;
    \draw (f) -- (h) ;
    \draw (g) -- (e) ;
    \node[label={[yshift=-0.5em, text width=10em, align=center]Integrated Live Video (Interest~$\sim P,\pi_0$)}] at (m) {};
    \node [below left=2.5em and 6em of m] (v1) {};
    \node [below left=2.5em and 0em of m] (v2) {};
    \node [draw, below of=m, node distance=2.8em,minimum width=9em] (vd) {Live Viewers};
    \node [below right=2.5em and 6em of m] (vN) {};
    \draw[->] (m) to (vd);
    \node (v1b)[left of=v1, node distance=3em] {};
    \node (vNb)[right of=vN, node distance=3em] {};
    \draw[dashed,->] (v1b.west) |-  node[left, rotate=90, yshift=0.5em, xshift=-0.75em] {Viewer Engagement} (a.west);
    \draw[dashed,->] (vNb.east) |- (a.east);
    \draw[dashed] (v1b.west) to (v1.west);
    \draw[dashed] (vd.west) to (v1.west);
    \draw[dashed] (vNb.east) to (vN.east);
    \draw[dashed] (vd.east) to (vN.east);
\end{tikzpicture}
\caption{Block diagram showing the stochastic scheduling problem faced by the decision maker (broadcaster) in advertisement scheduling on live media. The setup is detailed in Section~\ref{sec:numerical:results} of the paper. The broadcaster wishes to schedule at most $L$-ads during the live session. To maximize advertisement revenue, the ads need to be scheduled when the interest in the content is high. The interest in the content cannot be measured directly, but noisy observations of the interest are obtained from the viewer engagement (viewer comments and likes) during the live session.}
\label{fig:blkdiagram:live:scheduling}
\end{figure}
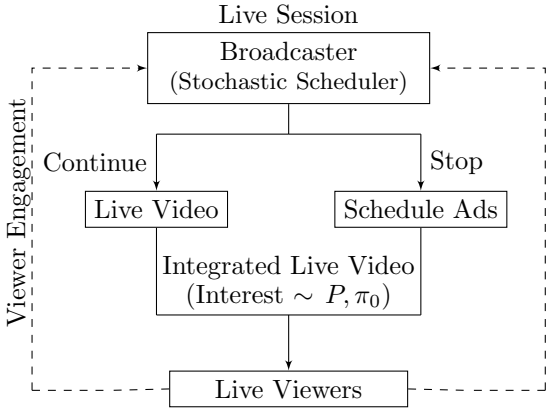

In this paper, motivated by the problem of interactive advertisement~{(ad)} scheduling in personalized live social media, we consider a {\em multiple stopping time problem} in a partially observed Markov chain. Figure~\ref{fig:blkdiagram:live:scheduling} shows the schematic setup of the ad scheduling problem considered in this paper. 
The broadcaster (decision maker) in Figure~\ref{fig:blkdiagram:live:scheduling} wishes to schedule \emph{at most} $L$ ads to maximize the cumulative advertisement revenue. 

\subsection*{Main results and Organization}
The multiple stopping time problem considered in this paper is a non-trivial generalization of the single stopping time problem, in that applying the single stopping policy multiple times doesn't yield the maximum possible cumulative reward; see Section~\ref{sec:numerical:results} for a numerical example. 
Section~\ref{sec:system:model} formulates the stochastic control problem faced by the decision maker (Broadcaster in Figure~\ref{fig:blkdiagram:live:scheduling}) as a multiple stopping time partially observed Markov decision process (POMDP); the POMDP formulation is natural in the context of a partially observed multi-state Markov chain with multiple actions ($L$ stops, continue). It is well known that for a POMDP, the computation of the optimal policy is PSPACE-complete~\cite{krishnamurthy2016partially}.  
Hence, we provide structural results on the optimal multiple stopping policy. 
The structural results are obtained by imposing sufficient conditions on the model - the main tools used are submodularity and stochastic dominance on the belief space of posterior distributions. 

This paper has the following main results:

{\em 1. Optimality of threshold policies: }Section~\ref{subsec:main:result} provides the main structural result of the paper. 
Specifically, Theorem~\ref{thm:main} asserts that the optimal policy is characterized by up to $L$ threshold curves, $\Gamma_l$ on the unit simplex of Bayesian posteriors (belief states). To prove this result we use the monotone likelihood ratio (MLR) stochastic order since it is preserved under conditional expectations. However, determining the optimal policy is non-trivial since the policy can only be characterized on a partially ordered set (more generally, a lattice) within the unit simplex. We modify the MLR stochastic order to operate on line segments within the unit simplex of posterior distributions. Such line segments form chains (totally ordered subsets of a partially ordered set) and permit us to prove that the optimal decision policy has a threshold structure. 
In addition, similar to~\cite{nak85}, we show that the stopping sets (set of belief states at which the decision maker stops) have a nested structure. 

{\em 2. Monotonicity of cumulative reward with transition matrix: } Section~\ref{subsec:copositivity} characterizes how the cumulative reward changes with respect to copositive ordering of the transition matrix. Specifically, Theorem~\ref{thm:copositive:reward} asserts that the optimal cumulative reward is monotone with respect to the copositive ordering of the transition matrix. The result can be used to implement reduced complexity posterior calculations for Markov chains with large dimension state space. 

{\em 3. Optimal Linear Threshold and their Estimation: }For the threshold curves $\Gamma_l, l=1,\cdots,L$, Theorem~\ref{thm:coefficent:MLR:condition} and Theorem~\ref{thm:constraints:parameter} give necessary and sufficient conditions for the optimal linear hyperplane approximation (linear threshold policies) that preserves the structure of the optimal multiple stopping policy. 
Section~\ref{sec:spsa:MLR:threshold} presents a simulation based stochastic gradient algorithm (Algorithm~\ref{algo:policy:gradient:algorithm}) to compute the optimal linear threshold policies. The advantage of the simulation based algorithm is that it is very easy to implement and is computationally efficient. 

{\em 4.~Application to Interactive Advertising in live social media: }
To illustrate the usefulness of the structural results for the multiple stopping time problem, we consider the application of interactive advertisement scheduling in personalized live social media.  
The problem of optimal scheduling of ads has been studied in the context of advertising in television; see~\cite{BBM04}, \cite{PC15} and the references therein. However, scheduling ads on live online social media is different from scheduling ads on television in two significant ways~\cite{KM11}: \begin{inparaenum}[i)]\item real-time measurement of viewer engagement (comments and likes on the content). The viewer engagement provides a noisy measurement of the underlying interest in the content. \item revenue is based on viewer engagement with the ads rather than a pre-negotiated contract\end{inparaenum}. 

In Section~\ref{sec:numerical:results}, we use a real dataset from Periscope, a popular personalized live streaming application owned by Twitter, to optimally schedule multiple ads ($L > 1$) in a sequential manner so as to maximize the advertising revenue. The numerical results show that the policy obtained through the multiple stopping framework outperforms conventional scheduling techniques. 
\subsection*{Context and Related Literature}
The problem of optimal multiple stopping has been well studied in the literature. In the classic $L$-secretary problem~\cite{Kle05}, independent and identically (i.i.d) observations are presented sequentially to the decision maker and the objective is to select $L$ observations so as to maximize the sum of reward (a function of observation). The classical setting with i.i.d observations have been extended to consider variety of scenarios such as the observation times arising out of Poisson process~\cite{sta87}, observations with a joint distribution and possibly depending on the stopping times in~\cite{Nik99} and for random horizon in~\cite{Ann15}. 
However, very few works consider optimal multiple stopping over a partially observed Markov chain. 
The closest work is due to Nakai~\cite{nak85} who considers optimal $L$-stopping over a finite horizon of length $N$ in a partially observed Markov chain. In~\cite{nak85}, properties of the value function and the nested property of the stopping regions is derived. The major weakness of~\cite{nak85} is the absence of an algorithm to compute the optimal policy utilizing the structural results. In addition, for many practical applications such as the interactive advertisement scheduling problem considered in this paper, the length of the horizon is not known apriori. Hence, this paper considers the multiple stopping problem over an infinite horizon, derives additional structural results compared to~\cite{nak85} and provides a stochastic gradient algorithm to compute optimal approximation policies satisfying the structural results.    

The optimal multiple stopping time problem can be contrasted to the recent work on sequential sampling with ``causality constraints''. 
\cite{ER15} considers the case where a decision maker is limited to a finite number of observations (sampling constraints) and must adaptively decide the observation strategy so as to perform quickest detection on a data stream. The extension to the case where the sampling constraints are replenished randomly is considered in~\cite{GBL14}. In the multiple stopping time problem, considered in this paper, there is no constraint on the observations and the objective is to stop at most $L$ times to maximize the cumulative reward. 

The optimal multiple stopping time problem, considered in this paper, is similar to the sequential scheduling problem with uncertainty~\cite{NJ10} and the optimal search problem considered in the literature. \cite{ST05} and \cite{LPVZ15} consider the problem of finding the optimal launch times for a firm under strategic consumers and competition from other firms to maximize profit. However, in this paper, we deal with sequential scheduling in a partially observed case.  \cite{WRA11} consider an optimal search problem where the searcher receives imperfect information on a (static) target location and decides optimally to search or interdict by solving a classical optimal stopping problem ($L=1$). However, the multiple-stopping problem considered in this paper is equivalent to a search problem where the underlying process is evolving (Markovian) and the searcher needs to optimally stop $L>1$ times to achieve a specific objective. 

Apart from interactive advertising, there are several other applications of the multiple stopping problem considered in this paper: American options with multiple exercise times~\cite{CT08}, $L$-commodities problem~\cite{sta87}, investment decision making~\cite{DL15}, to name a few.  

\section{Sequential multiple stopping and Stochastic dynamic programming}
\label{sec:system:model}
In this section, we formulate the optimal multiple stopping time problem as a POMDP. In Section~\ref{subsec:problem:formulation:stochastic:dynamic:programming}, we present a solution to the POMDP using stochastic dynamic programming. 
This sets the stage for Section~\ref{sec:scheduling:policy} where we analyze the structure of the optimal policy. 
\subsection{Optimal Multiple Stopping: POMDP Formulation}
\label{subsec:ad:scheduling:problem}
	Consider a discrete time Markov chain $\Root_\Time$ with state-space $\statespace = \left\{1,2, \cdots, {\statedim} \right\}$. 
	Here, $\Time = 0,1,\cdots$ denote discrete time. The decision maker receives a noisy observation $\obs_\Time$ of the state $\Root_\Time$ at each time $\Time$. The decision maker wishes to stop at most $L$ times over an infinite horizon. The positive integer $L$, is chosen a priori\footnote{The number of stops is a design parameter available to the decision maker. In this paper, we do not consider the problem of choosing the optimal number of stops. }.  
At each time the decision maker either stops or continues, and obtains a reward that depends on the current state of the Markov chain. 
The objective of the decision maker is to opportunistically select the best time instants to stop so as to maximize the cumulative reward. 
This problem of stopping at most $L$ times sequentially so as to maximize the cumulative reward corresponds to a multiple stopping time problem with $L$-stops. 

The multiple stopping time problem consists of the following components:

{\em 1.~State Dynamics:}
The Markov chain has transition matrix $\tp$ and initial probability vector $\belief_0$; so
\begin{equation}
	\label{eq:rootdynamics}
	\tp(i,j) = \prob(\Root_{\Time+1} = j|\Root_\Time = i),  \;  \belief_0(i) = \prob(\Root_0 = i).
\end{equation}

{\em 2.~Observations:}
At each time instant $t$, the decision maker receives noisy observation $\obs_\Time$ of the state $\Root_\Time$. Denote, the conditional probability of receiving observation $j \in \obspace$ ($\obs_t=j$) in state $i$ ($X_t=i$) by $B(i,j)$. Then, 
\begin{equation}
	\label{eq:obprob}
		\obp(i,j) = \prob\left(\Obs_{\Time} = j|\state_\Time = {i} \right) \; \forall i \in \statespace,j \in \obspace. 
\end{equation} 

{\em 3.~Actions:}
At each time instant $t$, the decision maker chooses an action $\action_\Time \in \actionspace=\{1\text{ (Stop) },2 \text{ (Continue) }\}$ to either stop or to continue. 

{\em 4.~Reward:}
Choosing the stop action at time $\Time$, when there are $l$ additional stops remaining, the decision maker accrues a reward\footnote{In the interactive advertisement scheduling application, the reward is indexed by the number of stops remaining to denote the varying ad revenue from the different ads placed during a session. \label{footnote:l:explanation}} $r_l(\Root_t,a=1)$, where $\Root_t$ is the state of the Markov chain at time $t$. Similarly, if the decision maker chooses to continue, it will accrue $r_l(\Root_t,a=2)$. 

{\em 5.~Scheduling Policy:} 
The \emph{history} available to the decision maker at time $t$ is
	\begin{align*}
		\history_\Time = \left\{\belief_0,\action_0, \obs_{1}, \cdots, \action_{\Time-1},\obs_{\Time} \right\}. 
	\end{align*}
The scheduling policy $\policy$, at each time $t$, maps $\history_\Time$ to action $\action_\Time$ i.e.\ the action chosen at time $\Time$ is $\action_\Time = \policy(\history_\Time)$. Let $\policyspace$ denote the set of admissible policies. 	
\textbf{Objective:}

For $l \in \left\{1,2,\cdots, L\right\}$, let $\tau_l$ denote the stopping time when there are $l$ stops remaining, i.e.\   
\begin{equation}
	\tau_l = \inf \left\{t: t > \tau_{l+1}, \action_t = 1 \right\}, \text{with } \tau_{L+1} = 0.
	\label{eqn:stopping:time:l}
\end{equation}

For policy $\policy$ and initial belief $\belief_0$, the cumulative reward is: 
\begin{align}
		\discountedreward_{{\policy}}(\belief_0)  	&= \E_\mu\left\{\sum_{\Time=0}^{\tau_L-1}\discount^{\Time}r_L(\state_\Time,2) + \discount^{\tau_L}r_L(\state_{\tau_L},1) \right. \label{eq:discountedreward_h} \\
							 & \left. + \sum_{\Time={\tau_L+1}}^{\tau_{L-1}-1}\discount^{\Time}r_{L-1}(\state_\Time,2)+ \dots +    \discount^{\tau_1}r_1(\state_{\tau_1},1) \given[\Big] \pi_0 \right\}, \nonumber  
	\end{align}
where the expectation is over the state dynamics and the observation distribution. 
In~\eqref{eq:discountedreward_h}, $\rho\in \left[0,1\right]$ denotes a user-defined economic discount factor\footnote{In the multiple stopping time problem, considered here, $\rho = 1$ is allowed. For undiscounted problem ($\rho = 1$), the stopping times may not be finite and the objective in~\eqref{eq:discountedreward_h} becomes unbounded. However, the multiple stopping time problem considered in this paper will terminate in finite time: Assume  $\overline{R} = \underset{i,l}{\max\;} r_l(i,1) > 0$ i.e.\ the maximum stop reward is positive and $\munderbar{R} = \underset{i,l}{\min \;} r_l(i,2) < 0$, i.e.\ the minimum reward to continue is negative. Then, it is clear that any optimal policy will stop in less than $\bar{T} = \cfrac{L \overline{R}}{|{\munderbar{R}}|}$ time steps. The intuition is that if $T > \bar{T}$ then the accumulated reward is negative and can be strictly improved by taking a stop action before $\bar{T}$. \label{foot:undiscount}}. 
Choosing $\rho < 1$ de-emphasizes the effect of decisions taken at later time instants on the cumulative reward. 
	
The decision maker aims to compute the optimal strategy $\optpolicy$ to maximize~\eqref{eq:discountedreward_h}, i.e.\  
\begin{equation}
	\optpolicy = \underset{\policy \in \policyspace}{\argmax} \; \discountedreward_{{\policy}}(\belief_0).  
	\label{eqn:opt:policy:definition}
\end{equation}
\begin{remark}\normalfont
The above formulation is an instance of a special type of POMDP called the stopping time POMDP. 
This is seen as follows: the objective in~\eqref{eq:discountedreward_h} can be expressed as an infinite horizon criteria by augmenting a fictitious absorbing state--$0$ that has zero reward, i.e.\ $r_0(0,\action) = 0 \; \action \in \actionspace$. When $L$ stop actions are taken, the system transitions to state $0$ and remains there indefinitely. Then~\eqref{eq:discountedreward_h} is equivalent to the following discounted infinite horizon criteria:
\begin{align*}
	\discountedreward_{{\policy}}(\belief_0)  	&= \E_\mu\left\{\sum_{\Time=0}^{\tau_L-1}\discount^{\Time}r_L(\state_\Time,2) + \discount^{\tau_L}r_L(\state_{\tau_L},1) \right. \\
							 & \left. + \dots +    \discount^{\tau_1}r_1(\state_{\tau_1},1)  + \sum_{\Time={\tau_1+1}}^{\infty}\discount^{\Time}r_{0}(0,2) \given[\Big] \pi_0 \right\},
\end{align*}
where the last summation is zero.
\end{remark}
\begin{remark}[Finite horizon constraint]\normalfont
This paper considers the problem of at most $L$ stops with no constraints on the stopping times. 
Our results also hold straightforwardly for the case where $L$ stops need to be made within a pre-specified finite time horizon. 
Then, the optimal policy will be non-stationary and the structural results presented in subsequent sections apply at each time instant. 
\end{remark}
\subsection{Belief State Formulation of the Objective}
As is customary for partially observed control problems, we reformulate the dynamics and cumulative objective in terms of the belief state. 
Let $\Belief$ denote the belief space of $S$-dimensional probability vectors. The belief space is the unit $S-1$ dimensional simplex: 
\begin{equation}
	\Belief = \left\{\belief : 0 \le \belief(i) \le 1, \sum_{i=1}^S \belief(i) = 1\right\}.
\end{equation}
The belief state at time $\Time$, denoted by $\belief_\Time \in \Belief$, is the posterior probability of $X_\Time$ given the history $\history_\Time$. The belief state  is a sufficient statistic of $\history_\Time$~\cite{bertsekas1995dynamic}, and evolves according to the following Hidden Markov Bayesian filter update~\cite{krishnamurthy2016partially}: 
\begin{equation} 
	\label{eq:hmmfilter}
	\begin{aligned}
		& \belief_{\Time+1} = \filter(\belief_{\Time},\obs_{\Time+1}), \quad \text{where}\\
		&\filter(\belief,\obsy) = \cfrac{\obp_{\obsy}\tp^{'}\belief}{\filternorm(\belief,\obsy)}, \quad \filternorm(\belief,\obsy) = \one_{\statedim}^{\prime}\obp_{\obsy}\tp^{'}\belief, \\
&\obp_{\obsy} = \text{diag}\left(\obp(1,\obsy), \cdots, \obp(\statedim,\obsy)\right).
\end{aligned}
\end{equation}
Here $\one_\statedim$ represents the $\statedim$-dimensional vectors of ones. 

Using the smoothing property of conditional expectations, the objective in~\eqref{eq:discountedreward_h} can be reformulated in terms of belief state as: 
\begin{align}
			& \discountedreward_{{\policy}}(\belief_0) 	= \E_\mu\left\{\sum_{\Time=0}^{\tau_L-1}\discount^{\Time}r_{2,L}^\prime \belief_{\Time}  + \discount^{\tau_L}r_{1,L}^\prime \belief_{\tau_L} \right. \label{eq:discountedreward} \\
				& \left. + \sum_{\Time={\tau_L+1}}^{\tau_{L-1}-1}\discount^{\Time}r_{2,L-1}^\prime \belief_{\Time} + \dots +    \discount^{\tau_1}r_{1,1}^\prime \belief_{\tau_1} + \sum_{\Time={\tau_1+1}}^{\infty}\discount^{\Time}r_{2,0}^\prime \belief_{\Time} \given[\Big] \pi_0 \right\}, \nonumber
\end{align}
where $r_{\action, l} = \left[r_l(1,\action), \dots, r_l(\statedim,\action)\right]^\prime$. 
For the stopping time problem~\eqref{eq:discountedreward}, there exists a stationary optimal policy~\cite{bertsekas1995dynamic}. Since the belief state is a sufficient statistic of $\history_\Time$, \eqref{eqn:opt:policy:definition} is equivalent to computing the optimal stationary policy $\optpolicy: \Belief \times \left[L\right] \rightarrow \actionspace$, where $\left[L\right] = \left\{1,2,\cdots,L\right\}$, as a function of belief and number of stops remaining to maximize~\eqref{eq:discountedreward}. 

\subsection{Stochastic Dynamic Programming}
\label{subsec:problem:formulation:stochastic:dynamic:programming}
Computing the optimal policy $\optpolicy$ to maximize~\eqref{eqn:opt:policy:definition} or equivalently~\eqref{eq:discountedreward} involves solving multiple stopping Bellman's dynamic programming equation~\cite{bertsekas1995dynamic}: 
\begin{equation}
\label{eq:bellman}
\begin{aligned}
\optpolicy(\belief,\looks) &= \underset{\action \in \actionspace}{\argmax} ~ Q(\belief,\looks,\action), \\
\valuefunction(\belief,\looks) &= \underset{\action \in \actionspace}{\max} ~ Q(\belief,\looks,\action),  
\end{aligned}
\end{equation}
\begin{equation*}
	\begin{aligned}
		Q(\belief,\looks, 1) 	&=    r_{1,l}^\prime\belief + \discount\sum_{y \in \obspace}\valuefunction\left(\filter(\belief,y),\looks-1\right)\filternorm(\belief,y), \\
		Q(\belief,\looks, 2) 	&=    r_{2,l}^\prime\belief + \discount\sum_{y \in \obspace}\valuefunction\left(\filter(\belief,y),\looks\right)\filternorm(\belief,y).  
	\end{aligned}
	\end{equation*} 
{\em \underline{Discussion}: } 
In~\eqref{eq:bellman}, $V(\pi,l)$ denotes the optimal value function at belief $\pi$ when $l$ stops are remaining, and is the expected accumulated reward induced by the optimal policy $\mu^*$. The optimal value function is the fixed point solution of the set of Bellman equations in~\eqref{eq:bellman}. The fixed point solution can be obtained using the value iteration algorithm (see~\ref{subsec:value:iteration}). $Q(\pi,l,u)$ is the expected accumulated reward starting at belief $\pi$ when $l$ stops remaining, and taking action $u$ and then using the optimal policy $\mu^*$. The Bellman equations can be explained as follows: When a stop action ($u = 1$) is taken, the decision maker obtains an instantaneous reward $r_{1,l}^\prime\belief$ and the number of stops remaining reduce by $1$. When the continue action is taken ($u = 2$), the decision maker obtains an instantaneous reward of $r_{2,l}^\prime\belief$, and the number of stops remaining is unaffected. The belief evolves according to~\eqref{eq:hmmfilter}. The second term in the summation computes the expected future reward where the expectations is with respect to the observation distribution. 

Since the state-space $\Belief$ is a continuum, Bellman's equation~\eqref{eq:bellman} or the value iteration algorithm in~\ref{subsec:value:iteration} does not translate into a practical solution methodology as $\valuefunction(\belief,\looks)$ needs to be evaluated at each $\belief \in \Belief$.  This, in turn, renders the computation of the optimal policy $\optpolicy(\belief,\looks)$ intractable\footnote{It is well known that a finite horizon POMDP with finite observation space can be solved exactly, indeed the value function is piecewise linear and convex~\cite{krishnamurthy2016partially}. However, the problem is PSPACE complete~\cite{PT87}; the worst case computational cost increases exponentially with the number of actions and doubly exponential with the time index.}. 
\section{Optimal Multiple Stopping: Structural results }
\label{sec:scheduling:policy}
In this section, we derive structural results for the optimal policy~\eqref{eq:bellman} of the multiple stopping time problem.  
In Section~\ref{subsec:main:result}, we show that under reasonable conditions on the POMDP parameters, the optimal policy is a monotone policy. 
In addition, in Section~\ref{subsec:copositivity}, we show the monotone property of the cumulative reward. 
\subsection{Definitions}
\label{subsec:definitions}
Define the stopping set $S^l$ (the set of belief states where Stop is the optimal action), when $l$ stops are remaining as: 
\begin{equation}
	S^l = \left\{\pi: \optpolicy(\pi,l) = 1\right\}. 
	\label{eqn:defn:stopping:region}
\end{equation}
Correspondingly, the continue set (the set of belief states where Continue is the optimal action) is defined as
\begin{equation}
	C^l = \left\{\pi: \optpolicy(\pi,l) = 2\right\}. 
	\label{eqn:defn:continue:region}
\end{equation}
Let $W(\pi,l)$ be defined as 
\begin{equation}
	W(\pi,l) = V(\pi,l)-V(\pi,l-1).
	\label{eqn:def:W}
\end{equation}
The stopping and continue sets in terms of $W$ defined in~\eqref{eqn:def:W} is as follows:
\begin{align}
	\begin{aligned}
		{S^{l}} = \{ \pi | r_l^\prime \pi \geq \rho \sum_y W(\Tpiy,l)\Spiy \},  \\
		{C^{l}} = \{ \pi | r_l^\prime \pi < \rho \sum_y W(\Tpiy,l)\Spiy \}.	
	\end{aligned}
\label{eqn:stop:continue:W}
\end{align}
where, $r_l \triangleq r_{1,l} - r_{2,l}$. 
\begin{remark}\normalfont
	For notational convenience, in this paper, without loss of generality, assume $r_{1,l} = r_l$ and $r_{2,l} = 0$. 
	So, the decision maker accrues no reward for the continue action. 
	Similarly, we consider $r_{1}=r_{2}=\cdots=r_{L} = r$, i.e.\ the rewards are not dependent on $l$. 
	It should be noted however that the structural results continue to hold for the case where the instantaneous rewards $r_l$ are dependent on $l$. 
\end{remark}

In general, the stopping and continue sets can be arbitrary partitions of the simplex $\Pi$. However, in Section~\ref{subsec:main:result}, we give sufficient conditions on the model so that these sets can be characterized by threshold curves. The question of computing the optimal policy, then, reduces to estimating the threshold curves. 

It is worth pointing out that in the classical stopping POMDPs in~\cite{krishnamurthy2016partially} with a single stop action, the stopping and continue sets are characterized in terms of convex value function. 
The key difficulty of the multiple stopping problem, considered in this paper, is that $W$ being the difference of two convex value functions does not share the convex properties of the value function. 

\subsection{Assumptions}
\label{subsubsec:assumptions}
The main result below, namely, Theorem~\ref{thm:main}, requires the following assumptions on the reward vector, $r$, the transition matrix, $P$ and the observation distribution, $B$. 
\begin{itemize}
	\item[\mylabel{ass:transition}{(A1)}] $P$ is totally positive of order $2$ (TP2), i.e.\ all second order minors are non-negative (see Definition~\ref{def:TP2:ordering} in Appendix~\ref{subsec:mlr:ordering}). 	
	\item[\mylabel{ass:observation}{(A2)}] $B$ is TP2. 
\item[\mylabel{ass:technical}{(A3)}] The vector, $\bar{r} = (I-\rho P)r$, has decreasing elements, i.e.\ $\bar{r}(1)\ge \cdots \ge\bar{r}(S)$.   
\end{itemize}

{\em Discussion of Assumptions: } 

\begin{inparaenum}

\item[]
		When $S=2$, \ref{ass:transition} is valid when $P(1,1) \ge P(2,1)$. When $S>2$, consider the tridiagonal transition matrix\footnote{The transition matrices computed on real dataset in Section~\ref{sec:numerical:results} follow a tridiagonal structure; refer to~\eqref{eqn:realdata:study}. } with $P(i,j) = 0, i>j+2 \text{ and } i<j-2$. \ref{ass:transition} is valid if $P(i,i) P(i+1,i+1) \ge P(i+1,i)P(i,i+1)$. 

\item[]
	\ref{ass:observation} holds for numerous examples. Examples include binomial, Poisson, geometric, Gaussian, exponential, etc. Table~{1.1}\footnotemark and Table~{1.2}\textsuperscript{\ref{footnote:muller}} in~\cite{mullercomparison} contains a detailed list.  
	In the numerical results in Section~\ref{sec:numerical:results}, we use the Poisson distribution where $B(i,j) = \frac{g_i^j \exp{(-g_i)}}{j!}$, where $g_i$ is the mean of the Poisson distribution. \ref{ass:observation} is satisfied if $g_i$ decreases monotonically with $i$. For a continuous observation distribution such as Gaussian whose mean is dependent on the state of the Markov chain (variance is fixed), \ref{ass:observation} is satisfied when the mean monotonically decreases with $i$. \footnotetext{The following continuous distribution satisfy \ref{ass:observation}: Exponential, Normal, Gamma, Weibull, Lognormal, Beta. Apart from numerous discrete probability mass functions, the following discrete distribution satisfy \ref{ass:observation}: Poisson, Binomial, Geometric.\label{footnote:muller}}

\item[]
	\ref{ass:technical} is a joint condition on the reward vector and the transition matrix. 
	Proposition~\ref{prop:r:decreasing:elements}, below, shows that \ref{ass:technical} and~\ref{ass:transition} jointly imply that the reward vector $r$ has decreasing elements. 
		When $S=2$, it can be verified that $r$ having decreasing elements is sufficient for $\left(I-\rho P\right)r$ to have decreasing elements. For $S > 2$, \ref{ass:technical} is a stronger condition than having the elements of $r$ decreasing.  
	
\ref{ass:technical} is easy to interpret when $P$ has additional structure. For example, consider a slowly varying Markov chain with $P = I + \epsilon Q$, where $Q(i,j)>0, i\ne j$, $\sum_j Q(i,j) = 0$, and $\epsilon >0$. 
Here $\frac{1}{\epsilon} > \underset{i}{\max}{\sum_j {|Q(i,j)|}}$ for $P$ to be a valid transition matrix. 
Then~\ref{ass:technical} is equivalent to $r$ having decreasing elements. Such slowly varying matrices arise in a lot of applications like manufacturing systems, internet packet transmission and wireless communication (see Section~{1.3} in~\cite{YZ06}). 
Also, the user interest in online social media typically evolves slowly~\cite{PB16}. 
The reward vector $r$ captures the preference of the decision maker - the highest reward is accrued in State $1$. 
\end{inparaenum}
\begin{proposition}
	If $P$ is TP2 and $(I-\rho P)r$ has decreasing elements, then $r$ has decreasing elements. 
	\label{prop:r:decreasing:elements}
\end{proposition}
The proof of Proposition~\ref{prop:r:decreasing:elements} is in Appendix~\ref{proof:prop:r:decreasing:elements}. 
\subsection{Main Result 1: Optimality of Threshold policies}
\label{subsec:main:result}
The main result below (Theorem~\ref{thm:main}) states that the optimal policy is monotone with respect to the belief state $\pi$. 
However, for a monotone policy to be well defined, we need to first define the ordering between two belief states. 
For $S=2$, the belief $\pi = \begin{bmatrix} 1-\pi(2)  &\pi(2)\end{bmatrix}$ can be completely ordered with respect to $\pi(2) \in \left[0,1\right]$. However, for $S>2$, comparing belief states requires using stochastic orders which are partial orders. 
We will use the monotone likelihood ratio (MLR) (see Def.~\ref{def:mlr:ordering} in Appendix~\ref{subsec:mlr:ordering}); it is ideal for partially observed control problems since it is preserved under conditional expectation (Bayesian update). 

Under reasonable conditions, Theorem~\ref{thm:main} asserts that the optimal policy $\mu^*(\pi)$ is monotonically decreasing in $\pi$ with respect to the MLR order. However, despite this monotonicity, determining the optimal policy is nontrivial since the policy can only be characterized on a partially ordered set. 
The main innovation in Theorem~\ref{thm:main} is to modify the MLR stochastic order to operate on \emph{lines} $\mathcal{L}(e_1,\bar{\pi})$ and $\mathcal{L}(e_S,\bar{\pi})$ (see~\ref{appendix:def:mlr:lines}) within the belief space. 
Such line segments form chains (totally ordered subsets of a partially ordered set) and permit us to prove that the optimal decision policy has a threshold structure.

\begin{figure}[h]
	\centering
	\scalebox{0.5}{\input{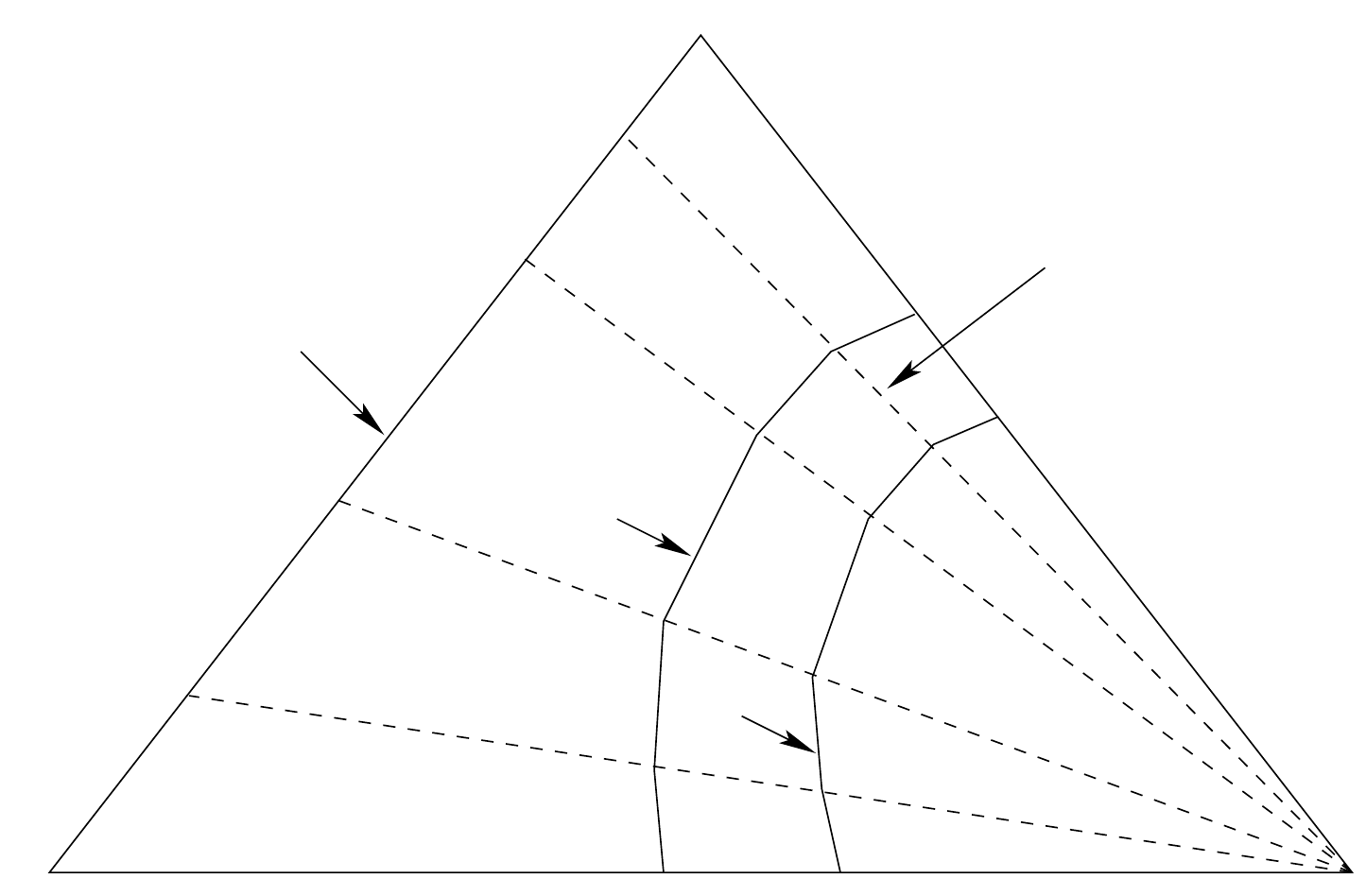}}
		\caption{Visual illustration of Theorem~\ref{thm:main}. Each of the stopping sets $S^l$ is characterized by a threshold curve $\Gamma_l$. Each of the threshold curve $\Gamma_l$ intersects the line $\mathcal{L}(e_1,\bar{\pi})$ at most once. }	\label{fig:mlr:lines}
\end{figure}
\begin{theorem}
	Assume {(A1)}, {(A2)} and {(A3)}. Then, 
	\begin{enumerate}[label=\Alph*]
		\item \label{thm:list:monotone:result}There exists an optimal policy $\mu^*(\pi,l)$ that is decreasing on lines $\mathcal{L}(e_1,\bar{\pi})$, and $\mathcal{L}(e_S,\bar{\pi})$ in the belief space $\Pi$ for each $l$\footnotemark.
		\item \label{thm:list:connected:sets}There exists an optimal switching curve $\Gamma_l$, for each $l$, that partitions the belief space $\Pi$ into two individually connected sets $S^l$ and $C^l$,such that the optimal policy is 
			\begin{equation}
				\mu^*(\pi,l) = 	\begin{cases}
					1 & \text{if } \pi \in S^l \\ 
					2 & \text{if } \pi \in C^l 
				\end{cases}
				\label{eqn:policy}
			\end{equation}
		\item \label{thm:list:subset}$S^{l-1} \subset S^l, l=1,2,\cdots,L$. 
	\end{enumerate}
	\label{thm:main}
\end{theorem}
\footnotetext{In general, the optimal policy is not unique. The theorem asserts that there exists a version of the optimal policy that is monotone. }
The proof of Theorem~\ref{thm:main} is given in Appendix~\ref{subsec:proof:thm:main}. \\
{\em \underline{Discussion}: }Theorem~\ref{thm:main}\ref{thm:list:monotone:result} asserts that the optimal policy is monotonically decreasing on the line $\mathcal{L}(e_1,\bar{\pi})$, as shown in Figure~\ref{fig:mlr:lines}. Hence, on each line $\mathcal{L}(e_1,\bar{\pi})$ there exists a threshold above (in MLR sense) which it is optimal to \emph{Stop} and below which it is optimal to \emph{Continue}. Theorem~\ref{thm:main}\ref{thm:list:connected:sets} asserts, for each $l$, the stopping and continue sets are connected. Hence, there exists a threshold curve, $\Gamma_l$, as shown in Figure~\ref{fig:mlr:lines}, obtained by joining the thresholds, from Theorem~\ref{thm:main}\ref{thm:list:monotone:result}, on each of the line $\mathcal{L}(e_1,\bar{\pi})$. 
Theorem~\ref{thm:main}\ref{thm:list:subset} proves the nested structure of the stopping sets: The stopping set when $l-1$ stops are remaining is a subset of the stopping set when there are $l$ stops remaining. 

In addition, Proposition~\ref{prop:stopping:set:convex:union}, below, shows that the stopping set enclosed by the threshold curve is a union of convex sets and hence, the threshold curve is continuous and differentiable almost everywhere. 
\begin{proposition}
	The stopping set $S^l$ is a finite union of convex sets. 
	\label{prop:stopping:set:convex:union}
\end{proposition}
The proof of Proposition~\ref{prop:stopping:set:convex:union} is in Appendix~\ref{proof:prop:stopping:set:convex:union}. 
\subsection{Main Result 2: Monotonicity of cumulative reward with transition matrix}
\label{subsec:copositivity}
Large transition matrices, common in real world applications, require large number of numerical computations to keep track of the belief dynamics in~\eqref{eq:hmmfilter}. 
Knowledge of the belief state is crucial to implement the optimal policy using a scheduler. 
One approach to deal with the computational bottleneck is to select a suitable transition matrix ``close'' to the true transition matrix such that the computation of the belief update is cheaper. 
It was shown in~\cite{KR14} that convex optimization techniques can be used to compute reduced rank matrices that bound (in terms of copositive ordering- Definition~\ref{def:copositive:ordering:transition:matrices} in Appendix) the true transition matrix $P$ from above and below, i.e.\ $\underbar{$P$} \preceq P \preceq \bar{P}$. 
Computing the belief state in~\eqref{eq:hmmfilter} requires $\mathcal{O}(S^2)$ computations, which could be expensive for large dimensional state space. 
The computational cost is reduced by using low rank (rank $R$) transition matrices ($\underbar{$P$}$ and $\bar{P}$)  which requires only $\mathcal{O}(RS)$ numerical operations. 
This leads us to the following question: How does the optimal cumulative reward of a multiple stopping time problem vary with transition matrix $P$? 
The main result below shows that if the transition matrices are partially ordered with respect to the copositive ordering so that $P \succeq \bar{P}$ then $J_{\mu^*(P)} \ge J_{\mu^*(\bar{P})}$.
\begin{theorem}
	Consider two multiple stopping time problems with transition matrices $P$ and $\bar{P}$, respectively, where $P \succeq \bar{P}$ with respect to copositive ordering (Definition~\ref{def:copositive:ordering:transition:matrices} in Appendix).
	If~\ref{ass:transition} to~\ref{ass:technical} hold, then the optimal cumulative rewards satisfy
	$$J_{\mu^*(P)} \ge J_{\mu^*(\bar{P})}.$$
	\label{thm:copositive:reward}
\end{theorem}
The proof of Theorem~\ref{thm:copositive:reward} is in Appendix~\ref{proof:copositive:theorem}; see also~\cite[Theorem~{14.8.1}]{krishnamurthy2016partially}.

{\em \underline{Discussion}: }Theorem~\ref{thm:copositive:reward} asserts that larger transition matrix (with respect to the copositive order) always results in a larger optimal reward.
This is useful in obtaining bounds on the achievable rewards in applications like interactive advertisement scheduling, where the interest dynamics change slowly over time. Also, the performance loss from using a low rank transition matrix for interest dynamics - to reduce the complexity of the real time scheduler - can be characterized. 

{\em \underline{Summary}: } This section derived the structural results of the optimal multiple stopping problem. 
The main structural result is in Theorem~\ref{thm:main}. Theorem~\ref{thm:main} generalizes the results in~\cite{nak85}. In addition to the nested property in~\cite{nak85}, Theorem~\ref{thm:main} characterizes the optimal policy by up to $L$ threshold curves. Additionally, Theorem~\ref{thm:copositive:reward} established the monotonicity of the optimal cumulative reward with respect to the copositive ordering of the transition matrix. 
\section{Stochastic Gradient Algorithm for estimating optimal linear threshold policies}
\label{sec:spsa:MLR:threshold}
In light of Theorem~\ref{thm:main}, computing the optimal policy reduces to estimating $L$-threshold curves in the unit simplex (belief space), one for each of the $L$-stops.  
The threshold curves can be approximated by any of the standard basis functions. 
In this paper, we will restrict the approximation to \emph{linear} threshold policies, i.e.\ policies of the form given in~\eqref{eqn:threshold:policy}. 
However, any such approximation needs to capture the essence of Theorem~\ref{thm:main}, i.e.\ the optimal policy is MLR decreasing on lines, connected and satisfy the nested property. 
We call such linear threshold policies (that captures the essence of Theorem~\ref{thm:main}) as the \emph{optimal} linear threshold policies. 

Section~\ref{subsec:structure:MLR:threshold} derives necessary and sufficient conditions to characterize such linear threshold policies. 
Algorithm~\ref{algo:policy:gradient:algorithm} in Section~\ref{subsec:spsa:linear:threshold:policy} is a simulation based algorithm to compute the optimal linear threshold policies. The simulation based algorithm is computationally efficient (see comments at end of Section~\ref{subsec:spsa:linear:threshold:policy}). 
\subsection{Structure of optimal linear threshold policies for multiple stopping}
\label{subsec:structure:MLR:threshold}
We define a linear parametrized policy on the belief space $\Pi$ as follows. 
Let $\theta_l \in \reals^{S-1}$ denote the parameters of linear hyperplane. 
Then, linear threshold policies as a function of the belief $\pi$ and the number of stops remaining $l$, are defined as 
\begin{equation}
	\mu_\theta(\pi, l) = 	
	\begin{cases}
		1 & \text{if } \begin{bmatrix} 0 & 1 & \theta_l \end{bmatrix} \begin{bmatrix} \pi \\ -1\end{bmatrix} \le 0 \\
		2 & \text{otherwise }.
	\end{cases}
	\label{eqn:threshold:policy}
\end{equation}
The linear policy $\mu_{\theta}(\pi,l)$ is indexed by $\theta$ to show the explicit dependence of the parameters on the policy. 
In~\eqref{eqn:threshold:policy}, $\theta = \left(\theta_1,\theta_2,\dots,\theta_L\right) \in \reals^{L \times (S-1)}$ is the concatenation of the $\theta_l$ vectors, one for each of the $L$-stops. \\
{\em \underline{Discussion}: }We will briefly discuss~\eqref{eqn:threshold:policy}: Given a general linear policy of the form $\alpha^\prime \pi \le \beta$, the specific form in~\eqref{eqn:threshold:policy} is obtained using \begin{inparaenum}[i)]\item the sum constraint on the belief $\belief$, i.e.\ $\sum_{i=1}^S \pi(i) = 1$, \item Scale invariance: For any positive constant $c$, $\alpha^\prime \pi \le \beta \Rightarrow c\alpha^\prime \pi \le c\beta$\end{inparaenum}. Also, notice that the dimension of both $\begin{bmatrix} 0 & 1 & \theta_l \end{bmatrix}$ and $\begin{bmatrix} \pi & -1\end{bmatrix}$ is $S+1$, since $\theta_l \in \reals^{S-1}$ and $\pi \in \reals^{S}$. 

In Theorem~\ref{thm:main}\ref{thm:list:monotone:result}, it was established that the optimal multiple stopping policy is MLR decreasing on specific lines within the belief space, i.e.\ for $\pi_1 \ge_{\mathcal{L}_i} \pi_2,\; \mu(\pi_1,l) \le \mu(\pi_2,l); i = 1,S$. 
Theorem~\ref{thm:coefficent:MLR:condition} gives necessary and sufficient conditions on the coefficient vector $\theta_l$ such that $\pi_1 \ge_{\mathcal{L}_i} \pi_2,\; \mu_{\theta}(\pi_1,l) \le \mu_{\theta}(\pi_2,l); i = 1,S$. 
\begin{theorem}
	A necessary and sufficient condition for the linear threshold policies $\mu_\theta(\pi,l)$ to be 
	\begin{enumerate}
		\item MLR decreasing on line $\mathcal{L}(e_1)$, iff $\theta_l(S-1) \ge 0$ and $\theta_l(i) \ge 0, \; i \le S-2$. 
		\item MLR decreasing on line $\mathcal{L}(e_S)$, iff $\theta_l(S-1) \ge 0$, $\theta_l(S-2) \ge 1$ and $\theta_l(i) \le \theta_l(S-2), \; i < S-2$. 
	\end{enumerate}
	\label{thm:coefficent:MLR:condition}
\end{theorem}
The proof of Theorem~\ref{thm:coefficent:MLR:condition} is in Appendix~\ref{appendix:proof:thm:lines:MLR:condition}. In Theorem~\ref{thm:coefficent:MLR:condition}, $\theta_l(i)$ denotes the $i^\text{th}$ element of the $S-1$ dimensional vector $\theta_l$.\\

{\em \underline{Discussion}: }As a consequence of Theorem~\ref{thm:coefficent:MLR:condition}, the constraints on the parameters $\theta$ ensure that only MLR decreasing linear threshold policies are considered; the necessity and sufficiency imply that non-monotone policies are not considered, and monotone policies are not left out. 

In Theorem~\ref{thm:main}\ref{thm:list:connected:sets} it was established that the optimal stopping sets are connected, which is satisfied trivially since we approximate the threshold curve using a linear hyperplane. Theorem~\ref{thm:constraints:parameter} below provides sufficient conditions such that the parametrized linear threshold curves satisfy the nested property established in Theorem~\ref{thm:main}\ref{thm:list:subset}. A proof is provided in Appendix~\ref{appendix:proof:thm:constraints:parameter}.

\begin{theorem}
		A sufficient condition for the linear threshold policies in~\eqref{eqn:threshold:policy} to satisfy the nested structure in Theorem~{\ref{thm:main}\ref{thm:list:subset}} is given by
	\begin{equation}
		\begin{aligned}
			\theta_{l-1}(S-1) &\le \theta_l(S-1) \\
			\theta_{l-1}(i) &\ge \theta_l(i) \quad i < S-1, 
		\end{aligned}
		\label{eqn:cond:parameter}
	\end{equation}
	for each $l$. 
	\label{thm:constraints:parameter}
\end{theorem}
\subsection{Simulation-based stochastic gradient algorithm for estimating linear threshold policies}
\label{subsec:spsa:linear:threshold:policy}
We now estimate the optimal linear threshold policies using a simulation based stochastic gradient algorithm using Algorithm~\ref{algo:policy:gradient:algorithm}. The algorithm is designed so that the estimated policies satisfy the conditions in Theorem~\ref{thm:coefficent:MLR:condition} and Theorem~\ref{thm:constraints:parameter}. 

The optimal policy of a multiple stopping time problem maximizes the expected cumulative reward $J_{\mu}$ in~\eqref{eq:discountedreward_h}. 
In Algorithm~\ref{algo:policy:gradient:algorithm}, we approximate $J_{\mu}$ over a finite time horizon ($N$)\footnote{For the optimal policy $\mu^*$, a horizon of length $N$ and the discount factor of $\rho$, $\lvert J_{\mu^*} - J_N \rvert_2 \le \frac{\rho^N}{1-\rho} \underset{l,x,u}{\max} \lvert r_l(x,u)\rvert$~\cite[Theorem~{7.6.3}]{krishnamurthy2016partially}. Hence, given an error tolerance $\varepsilon$, the horizon can be calculated as $N > \cfrac{\log\left(\frac{(1-\rho)\varepsilon}{\underset{l,x,u}{\max} \lvert r_l(x,u)\rvert}\right)}{\log \rho}$. \label{footnote:horizon}}, as $\discountedreward_N$ which is computed as: \begin{equation}
	\discountedreward_N(\theta) = {\E}_{\mu_\theta}\left\{\sum_{l=1}^{L}\discount^{\tau_l} r^\prime \pi_{\tau_l} \given[\Big] \tau_l \le N; \forall l\right\}. 
	\label{eqn:finite:time:approximation}
\end{equation}
$\discountedreward_N$ is an asymptotic estimate of $\discountedreward_\policy$ as $N$ tends to infinity.  
In~\eqref{eqn:finite:time:approximation}, we have made explicit the dependence of the parameter vector on the discounted reward and with an abuse of notation, have suppressed the dependence on the policy $\mu$. 

Algorithm~\ref{algo:policy:gradient:algorithm}, is a stochastic gradient algorithm that generates a sequence of estimates $\theta_n$, that converges to a local maximum. It requires the computation of the gradient: $\nabla_\theta J_N(\cdot)$. Evaluating the gradient in closed form is intractable due to the non-linear dependence of $\discountedreward_N(\theta)$ on $\theta$. 
We can estimate $\hat{\nabla}_\theta J_N(\cdot)$ using a simulation based gradient estimator. 
There are several such simulation based gradient estimators available in the literature including infinitesimal perturbation analysis, weak derivatives and likelihood ratio (score function) methods~\cite{pflug2012optimization}. In this paper, for simplicity, we use the SPSA algorithm~\cite{spall2005introduction}, which estimates the gradient using a finite difference method.  

To make use of the SPSA algorithm, we convert the constrained optimization problem in $\theta$ (constraints imposed by Theorem~\ref{thm:coefficent:MLR:condition} and Theorem~\ref{thm:constraints:parameter}) into an unconstrained problem using spherical co-ordinates as follows: \begin{equation}
	\theta^\phi_l(i) = 
	\begin{cases}
		\phi_1^2(S-1 ) \prod_{\ell=l}^{L-1} \sin^2(\phi_\ell(S-1)) & i = S-1\\
		1 + \phi_1^2(S-2) \prod_{\ell=2}^l \sin^2(\phi_\ell(S-2)) & i=S-2 \\
				\theta_l(S-2) \prod_{\ell = 1}^L \sin^2(\phi_\ell(i)) & i < S-2.
	\end{cases}
	\label{eqn:reparameterized:theta:phi}
\end{equation}
It can be verified that the parametrization, $\theta^\phi$ in~\eqref{eqn:reparameterized:theta:phi}, satisfies the conditions in Theorem~\ref{thm:coefficent:MLR:condition} and Theorem~\ref{thm:constraints:parameter}. For example, consider $i=S-1$, then the product term involving $\sin(\cdot)$ ensures that $\theta_{l-1}(S-1) \le \theta_l(S-1)$ (the first part of Theorem~\ref{thm:constraints:parameter}). 
\begin{algorithm}
\caption{Stochastic Gradient Algorithm for Optimal Multiple Stopping}
\label{algo:policy:gradient:algorithm}
\begin{algorithmic}[1]
       \Require POMDP parameters satisfy~\ref{ass:transition}, \ref{ass:observation}, \ref{ass:technical}. 
       \State Choose initial parameters ${\phi}_0$ and initial linear threshold policies $\mu_{{\theta}_0}$ using~\eqref{eqn:threshold:policy}. 
       \For{iterations $n=0,1,2,\dots$: }
              \State Evaluate $J_N(\theta^{{\phi}_n+c_n \omega_n})$ and $J_N(\theta^{{\phi}_n-c_n \omega_n})$ using~\eqref{eqn:finite:time:approximation} 
       \State SPSA: Gradient estimate $\hat{\nabla}_\phi J_N(\theta^{{\phi}_n})$ using~\eqref{eqn:spsa:gradient:estimate}. 
       \State Update the parameter vector ${{{\phi}_n}}$ to ${{{\phi}_{n+1}}}$ using~\eqref{eqn:spsa:inequality:parameter:update}. 
       \EndFor
\end{algorithmic}
\end{algorithm}

Following~\cite{spall2005introduction}, the gradient estimate using SPSA is obtained by picking a random direction $\omega_n$, at each iteration $n$. The estimate of the gradient is then given by
\begin{align}
	\hspace{-0.5em}\hat{\nabla}_\phi J_N(\theta^{{\phi}_n}) &= \frac{J_N(\theta^{{\phi}_n+c_n \omega_n}) - J_N(\theta^{{\phi}_n-c_n \omega_n})}{2c_n} \omega_n,\label{eqn:spsa:gradient:estimate}
	\shortintertext{where,}
	\omega_n(i) &= 	\begin{cases} 
				-1 & \text{with probability } 0.5 \\
				+1 & \text{with probability } 0.5. 
			\end{cases} \nonumber
\end{align}
The two $J_N(\cdot)$ terms in the numerator of~\eqref{eqn:spsa:gradient:estimate} is estimated using the finite time horizon approximation~\eqref{eqn:finite:time:approximation}. A more detailed description of the finite time horizon approximation in given in Algorithm~\ref{algo:JN} in~\ref{appendix:finite:time:approximation:algorithm}.  
Using the gradient estimate in~\eqref{eqn:spsa:gradient:estimate}, the parameter update is as follows~\cite{spall2005introduction}:
\begin{equation}
	\phi_{n+1} = \phi_n + a_n \hat{\nabla}_\phi J_N(\theta^{{\phi}_n}). 
	\label{eqn:spsa:inequality:parameter:update}
\end{equation}
The parameters $a_n$ and $c_n$ are typically chosen as follows~\cite{spall2005introduction}:
\begin{align}
	\begin{aligned}
		a_n &= \upvarepsilon(n+1+\varsigma)^{-\kappa} &0.5<\kappa\le 1, &  \; \text{and } \upvarepsilon,\varsigma>0\\
		c_n &= \upmu(n+1)^{-\upsilon}  & 0.5<\upsilon \le 1 &  \; \upmu > 0			\end{aligned}
	\label{spsa:parameters:an:cn:rn}
\end{align}
The decreasing step size stochastic gradient algorithm, Algorithm~\ref{algo:policy:gradient:algorithm}, converges to a local optimum with probability one. 
There are several methods available in the literature that can be used for stopping criteria in Step~2 of Algorithm~\ref{algo:policy:gradient:algorithm}~\cite{spall2005introduction}. In this paper, we used the following criteria: 
\begin{inparaenum}[(i)]
	\item Small gradient: $\lVert \hat{\nabla}_\phi J_N(\theta^{{\phi}_n}) \rVert_2 \le \varepsilon$. 
	\item Max Iteration: Iterations are stopped when a maximum number is reached. We use $1000$ as the maximum number of iterations in our simulations. 
\end{inparaenum}

At each iteration of Algorithm~\ref{algo:policy:gradient:algorithm}, evaluating the gradient estimate in~\eqref{eqn:spsa:gradient:estimate} requires two POMDP simulations. However, this is independent of the number of states, the number of observations or the number of stops.  

{\em \underline{Summary}:} We have used a stochastic gradient algorithm to estimate the optimal linear threshold policies for the multiple stopping time problem. 
Instead of linear threshold policies, one could also use piecewise linear or other basis function approximations; providing that the resulting parameterized policy is still MLR decreasing (i.e.\ the characterization similar to Theorem~\ref{thm:coefficent:MLR:condition} holds).  
\section{Numerical Examples: Interactive advertising in Live Social Media}
\label{sec:numerical:results}
This section has three parts. 
In Section~\ref{subsec:results:synthetic}, we illustrate the main result of the paper using numerical examples. 
Second, using a Periscope dataset, we study how the multiple stopping problem can be used to schedule advertisements in live social media. 
We show numerically that the linear threshold scheduling policies (derived in Sec.~\ref{sec:spsa:MLR:threshold}) outperforms conventional techniques for scheduling ads in live social media.  
Finally, we illustrate the performance of the linear threshold policies for a large size POMDP ($25$ states) by comparing with the popular SARSOP algorithm.
\subsection{Synthetic Data}
\label{subsec:results:synthetic}
This section has four parts. First, we visually illustrate the optimal multiple stopping policy, using numerical examples, for $S=3$\footnote{For $S=3$, the unit simplex is an equilateral triangle. }. 
The objective is to illustrate how the assumptions in Sec.~\ref{subsubsec:assumptions} affect the optimal multiple stopping time policy. 
The optimal policy can be obtained by solving the dynamic programming equations in~\eqref{eq:bellman} and can be computed approximately by discretizing the belief space. The belief space $\Pi$ , for all examples below, was uniformly quantized into $100$ states, using the finite grid approximation method in~\cite{LOV91}. 
Second, we illustrate how the optimal accumulated rewards varies with the number of stops. For Example~1, below, it is easy to see that the accumulated reward increase with $L$. 
Third, we benchmark the performance of linear threshold policies (obtained using Algorithm~\ref{algo:policy:gradient:algorithm}) against optimal multiple stopping policy. 
Finally, we illustrate the advantage of structural results for designing approximation algorithm by comparing the performance of the linear threshold policies in Sec.~\ref{sec:spsa:MLR:threshold} against the popular softmax parametrization, which are not constrained to satisfy the structural results. 

\textbf{Example 1:}
{\em POMDP parameters:} Consider a Markov chain with $3-$states with the transition matrix $P$ and the reward vector specified in~\eqref{eqn:pomdp:parameters}. 
The observation distribution is given by $B(i,j) = \frac{g_i^j \exp{(-g_i)}}{j!}$, i.e.\ the observation distribution is Poisson with state dependent mean vector $g$ given in~\eqref{eqn:pomdp:parameters}. 
It is easily verified that the transition matrix, the observation distribution and the reward vector satisfy the conditions~{(A1)} to~{(A3)}. 
\begin{gather}
	P = 
	\begin{bmatrix}
		0.2 & 0.1 & 0.7 \\
		0.1 & 0.1 & 0.8\\
		0 & 0.1 & 0.9 
	\end{bmatrix}, 
		g = 
	\begin{bmatrix}
		12 & 7  & 2
	\end{bmatrix}, 
		r = 
	\begin{bmatrix}
		9& 3& 1
	\end{bmatrix} 
		\label{eqn:pomdp:parameters}
\end{gather}
We choose\footnote{This is motivated by the real dataset example in Section~\ref{subsec:real:data:youtube:twitch}.} $L=5$, i.e.\ the decision maker wishes to stop at most $5$ times. 

\begin{figure*}[!bth]
\minipage{0.32\textwidth}
  \includegraphics[width=\linewidth]{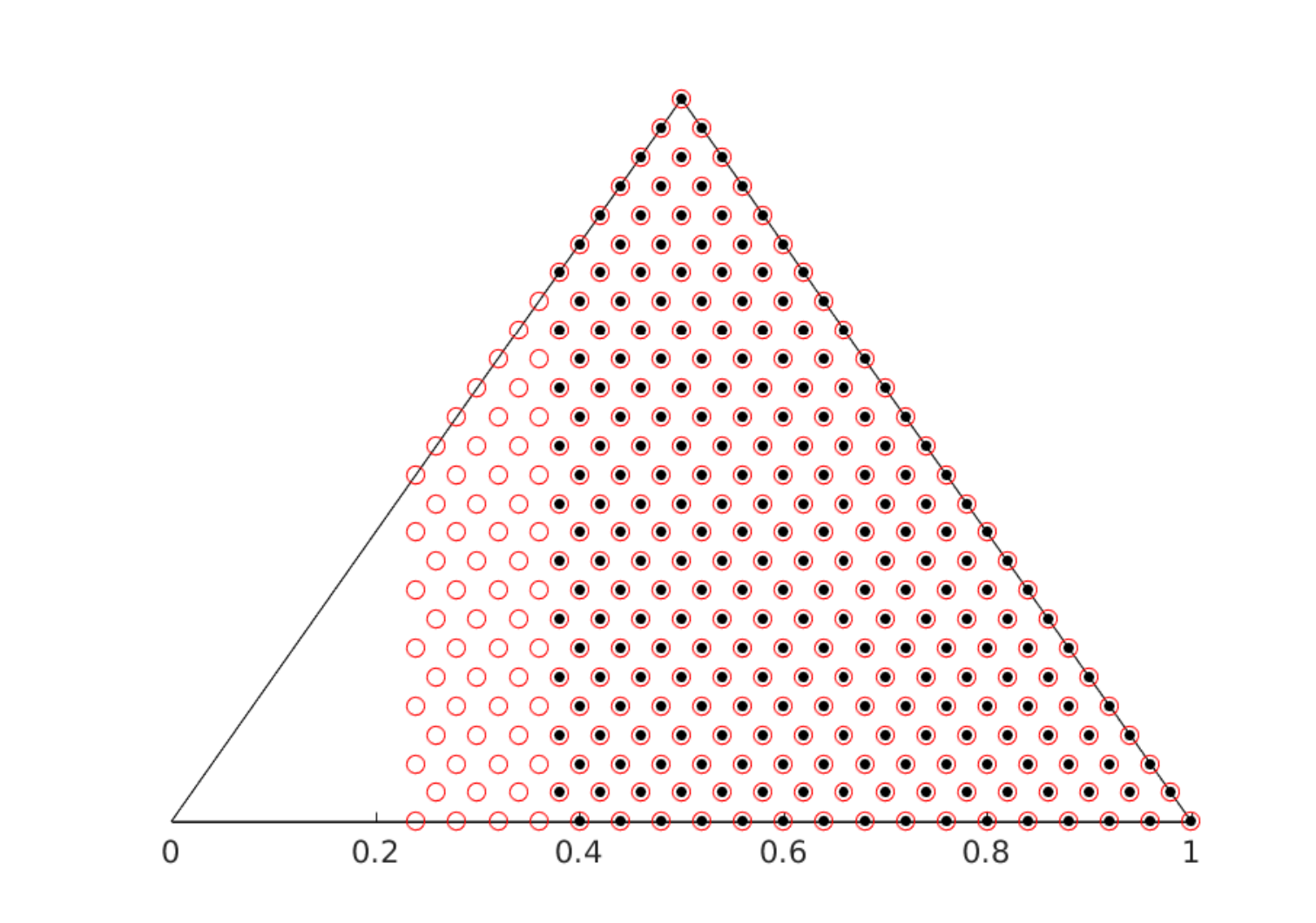}
  \caption{Example~1: $S^1$ (shown in black) and $S^5$ (shown in red) obtained by solving the dynamic programming~\eqref{eq:bellman}. The figure illustrates monotone, connected and the nested structure of the stopping sets ($S^{l-1} \subset S^l$), in Theorem~\ref{thm:main}. }\label{fig:simulation:study:stopping:set}
\endminipage\hfill
\minipage{0.32\textwidth}
  \includegraphics[width=\linewidth]{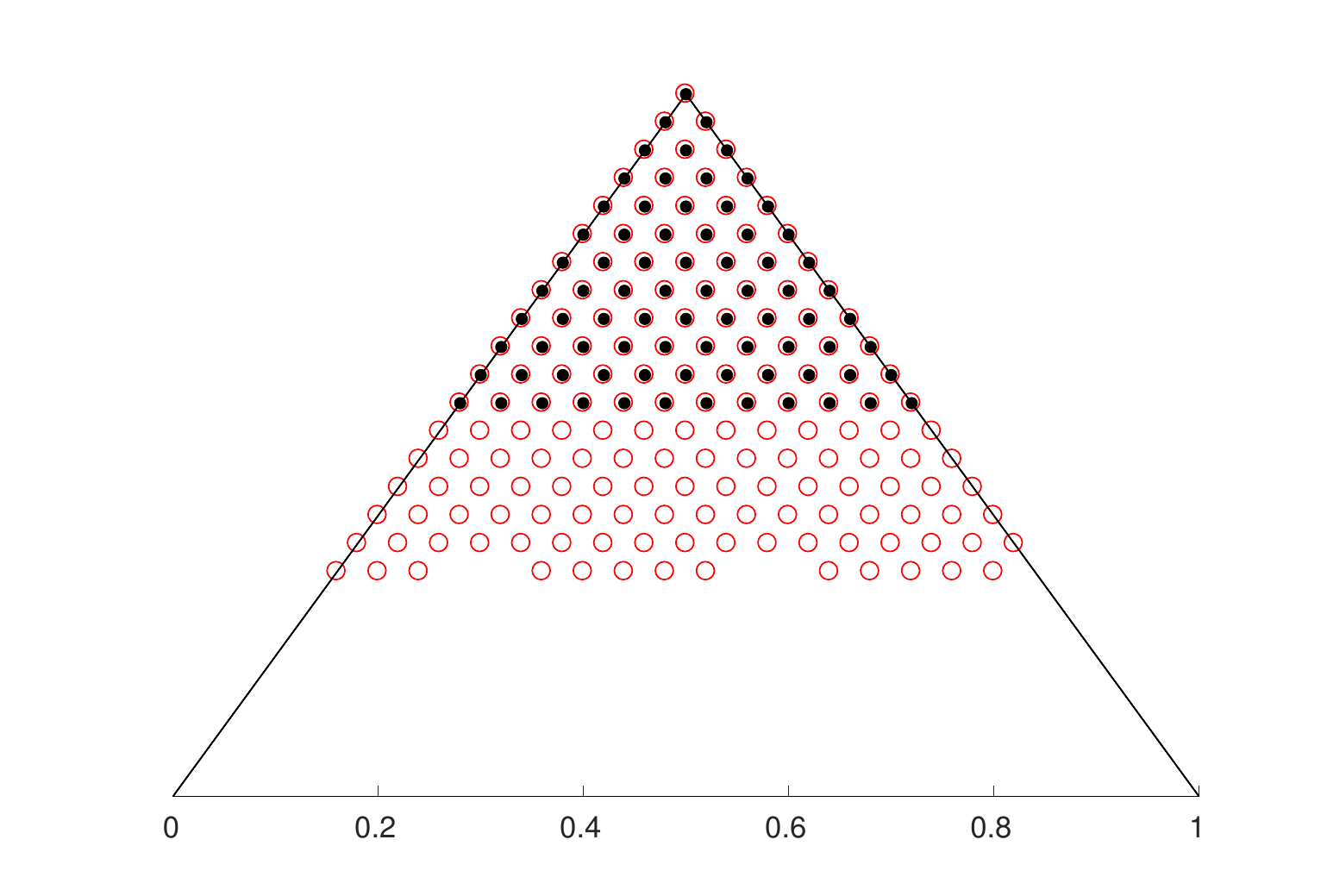}
  \caption{Example~2: Optimal policy when~\ref{ass:technical} is violated. $S^1$ is shown in black and $S^5$ is shown in red. The monotone property of Theorem~{\ref{thm:main}\ref{thm:list:monotone:result}} is violated. }\label{fig:A3:violated}
\endminipage\hfill
\minipage{0.32\textwidth}  \includegraphics[width=\linewidth]{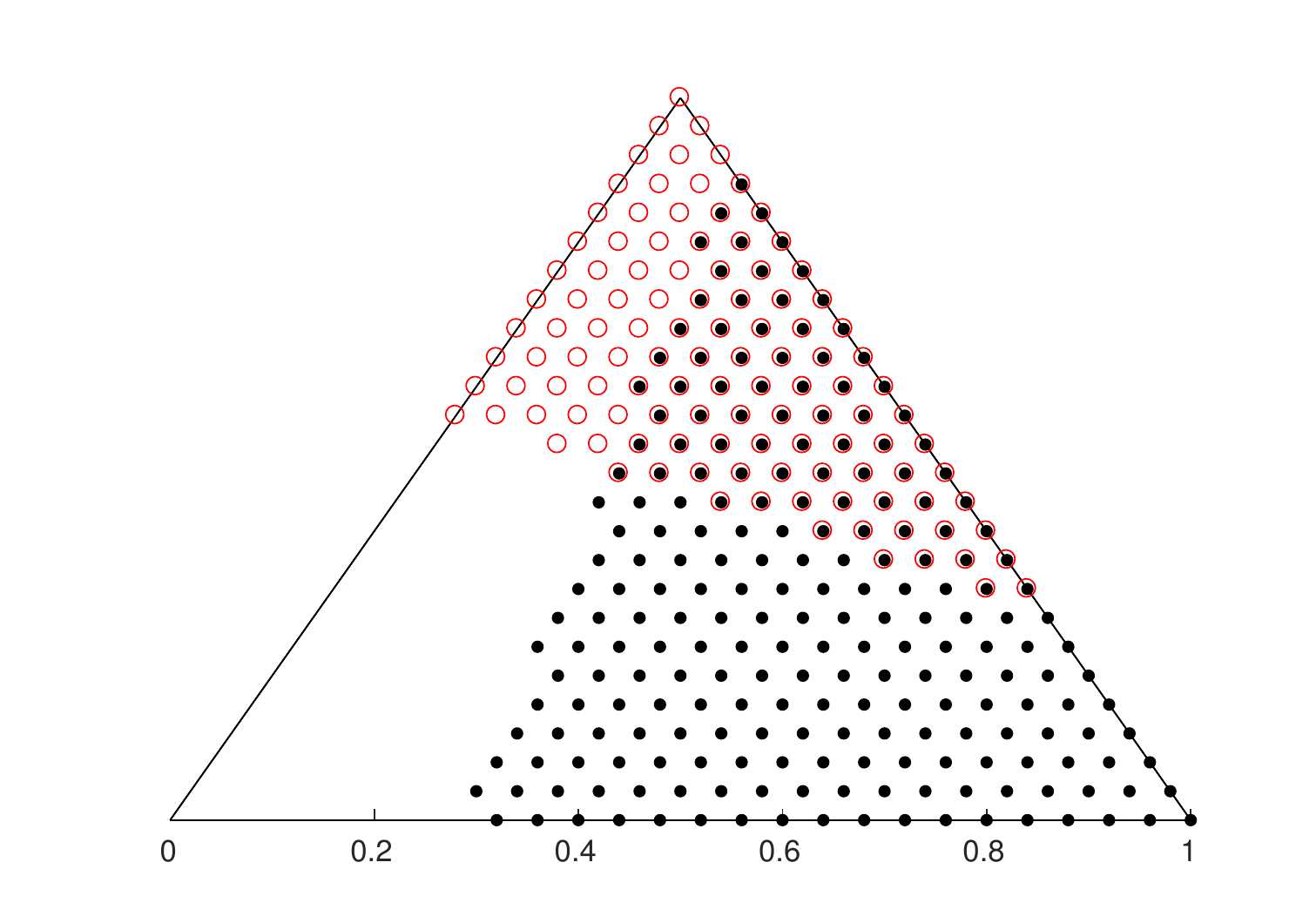}
  \caption{Example~3: Optimal policy when~\ref{ass:technical} is violated. $S^1$ is shown in black and $S^2$ is shown in red. The stopping sets are not nested. }\label{fig:A3:violated:L2}
\endminipage
\end{figure*}

Figure~\ref{fig:simulation:study:stopping:set} shows the stopping sets $S^5$ and $S^1$. 
It is evident from Figure~\ref{fig:simulation:study:stopping:set} that the optimal policy is monotone on lines, stopping sets are connected and satisfy the nested property; thereby illustrating Theorem~\ref{thm:main}. 

\textbf{Example 2:}
Consider the same parameters as in Example~1, except reward $r=\begin{bmatrix}1 & 2 & 1 \end{bmatrix}$ which violates ~\ref{ass:technical}. 
Figure~\ref{fig:A3:violated} shows the optimal multiple stopping policy in terms of the stopping sets. As can be seen from Figure~\ref{fig:A3:violated} that the optimal policy does not satisfy the monotone property (Theorem~{\ref{thm:main}\ref{thm:list:monotone:result}}). However, the nested property continues to hold.  

\textbf{Example 3:}
Consider the same parameters as in Example~1, except $L=2$ and $r_1=\begin{bmatrix}9 & 3 & 1 \end{bmatrix}$ and $r_2=\begin{bmatrix}3 & 9 & 1 \end{bmatrix}$. 
Assumption~\ref{ass:technical} is violated for $l=2$. 
Figure~\ref{fig:A3:violated:L2} shows the optimal multiple stopping policy in terms of the stopping sets. As can be seen from Figure~\ref{fig:A3:violated:L2} that the optimal policy does not satisfy the monotone property or the nested property. 

Thus, the conditions~\ref{ass:transition} to~\ref{ass:technical} of Theorem~\ref{thm:main} are useful in the sense that when they are violated, there are examples where the optimal policy does not have the monotone or nested property. 

{\em Optimal accumulated reward against $L$: }
Consider Example~1 with POMDP parameters in~\eqref{eqn:pomdp:parameters}. At each stop, we accumulate a reward. It is easy to see that as the number of stops increase, the reward accrued will also increase. Table~\ref{tab:cumulative:reward:L} illustrates that this is indeed the case. The values in Table~\ref{tab:cumulative:reward:L} were obtained by solving the dynamic programming equations in~\eqref{eq:bellman} for various values of $L$ ranging from $1-5$. \begin{table}[!h]
	\centering
	\caption{Optimal accumulated reward as a function of the number of stops. As the number of stops increase the accumulated reward increase. The table was generated by solving the dynamic programming equations in~\eqref{eq:bellman}. The accumulated reward is with a starting belief $\pi_0 = \left( \frac{1}{3} \; \frac{1}{3} \; \frac{1}{3} \right)$. }
	\label{tab:cumulative:reward:L}
	\begin{tabular}{|c|c|}
		\toprule
		$L$ 	& Cumulative reward\\
		 	& (Normalized w.r.t $L=1$)\\
		\midrule
		1 &  1.0000  \\
		2 &  1.6617  \\
		3 &  2.1154  \\
		4 &  2.4586  \\
		5 &  2.7519  \\
		\bottomrule
	\end{tabular}
\end{table}

{\em Performance of linear threshold policies: }
In order to benchmark the performance of optimal linear threshold policies (that satisfy the constraints in Theorem~\ref{thm:coefficent:MLR:condition} and Theorem~\ref{thm:constraints:parameter}), we ran Algorithm~\ref{algo:policy:gradient:algorithm} for Example~1 (parameters in~\eqref{eqn:pomdp:parameters}). 
The performance was compared based on the expected cumulative reward between the optimal policy and the linear threshold policies for $1000$ independent runs.  
The following parameters were chosen for the SPSA algorithm $\upmu = 2$, $\upsilon = 0.2$, $\varsigma=0.5$, $\kappa = 0.602$ and $\upvarepsilon = 0.1667$; these values are as suggested in~\cite{spall2005introduction}. 
It was observed that there is a $12\%$ drop in performance of the linear threshold policies compared to the optimal multiple stopping policy. 

{\em {{Advantage of parametrization satisfying structural results}}:} Here, we illustrate the advantage of parametrization of the policy to satisfy the structural results in Theorem~\ref{thm:main}. 
The softmax function is a popular parametrization for decision-making and is widely used in artificial neural networks~\cite{bishop2006pattern} and reinforcement learning~\cite{sutton1998reinforcement}.  
Consider the following softmax parametrization of the policy
\begin{equation}
	\Pr(\mu(\pi,l) = u) = \frac{\exp{\left(\begin{bmatrix} 0 & \theta_{l,u}\end{bmatrix}^\prime \pi\right)}}{\sum_{u=1}^2 \exp{\left(\begin{bmatrix} 0 & \theta_{l,u}\end{bmatrix}^\prime \pi\right)}}. 
	\label{eqn:soft-max}
\end{equation}
In~\eqref{eqn:soft-max}, $\Pr(\mu(\pi,l) = u)$ denotes the probability of taking action $u$ (either `Stop' or `Continue') as a function of belief $\pi$ and number of stops remaining $l$. The parameters in~\eqref{eqn:soft-max} $\theta_{l,u} \in \reals^{S-1}; l = 1,\cdots,L\; u=1,2$ are indexed by number of stops remaining and the actions. Compared to linear threshold policies in~\eqref{eqn:threshold:policy}, the policies in~\eqref{eqn:soft-max} are not restricted to satisfy the structural results in Theorem~\ref{thm:main}. 
Algorithm~\ref{algo:JN:softmax} in~\ref{appendix:finite:time:approximation:algorithm} summarizes the computation of the finite time horizon approximation with the softmax parametrization in~\eqref{eqn:soft-max}. 

Comparing the expected cumulative reward, we find that the optimal policy and the linear threshold policies outperform the softmax parametrization (Algorithm~\ref{algo:JN:softmax} in~\ref{appendix:finite:time:approximation:algorithm}) by $40\%$ and $30\%$, respectively. Hence, this illustrates the advantage of taking into account the structure of the optimal policy while designing algorithms for computing an approximation policy. 
\subsection{Real dataset: Interactive ad scheduling on Periscope using viewer engagement }
\label{subsec:real:data:youtube:twitch}
We now formulate the problem of interactive ad scheduling on {\emph{live}} online social media as a multiple stopping problem and illustrate the performance of linear threshold policies using a Periscope dataset\footnote{We use the dataset in~\cite{WZWZZ16}, which can be downloaded from \url{http://sandlab.cs.ucsb.edu/periscope/}. In~\cite{WZWZZ16}, the authors deal with the performance of Periscope application in terms of delay and scalability. }. 
Periscope is a popular live personalized video streaming application where a broadcaster interacts with the viewers via live videos. 
Each such interaction lasts between $10-20$ minutes and consists of: 
\begin{inparaenum}[(i)]
\item A broadcaster who starts a live video using a handheld device. 
\item Video viewers who engage with the live video through comments and likes. 
\end{inparaenum}

A strong motivation to consider the problem of interactive ad scheduling in live online videos stems from the fact that ads are currently scheduled using \textit{passive} techniques: periodic~\cite{SOW13}, and manual methods; and yet advertisement revenues are significant for social media companies\footnote{The revenue of Twitch which deals with live video gaming, play through of video games, and e-sport competitions, is around 3.8 billion for the year 2015, out of which 77\% of the revenue was generated from advertisements.}. The technique of interactive scheduling, where viewer engagement is utilized to schedule ads, has not been addressed in the literature. It will be seen in this section that interactive scheduling of ads has significant performance improvements over the existing passive methods. 

{\bf Dataset:}
The dataset in~\cite{WZWZZ16} contains details of all public broadcasts on the Periscope application from May~{15}, 2015 to August~{20}, 2015. 
The dataset consists of timestamped events: time instants at which the live video started/ended; time instants at which viewers join; and, time instants at which the viewers engage using \textit{likes} and \textit{comments}. 
In this paper, we consider viewer engagement through \textit{likes}, since \textit{comments} are restricted to the first $100$ viewers in the Periscope application. 

{\bf Ad scheduling Model}

Here we briefly describe how the model in Section~\ref{sec:system:model} can be adapted to the problem of interactive ad scheduling in live video streaming; see Figure~\ref{fig:blkdiagram:live:scheduling} for the schematic setup. 

{\em 1.~Interest Dynamics:} 
In live online social media, it is well known that the viewer engagement is correlated with the interest of the content being streamed or broadcast. 
Markov models have been used to model interest in online games~\cite{baldominos2016real}, web~\cite{AMM12} and in online social networks~\cite{BRTMA09}.
We therefore model the interest in live video as a Markov chain, $X_t$, where the different states denote the level of interest in the live content. 
The states are ordered in the decreasing order of interest. 

{\em Homogeneous Assumption: }Periscope utilizes the Twitter network to link broadcasters with the viewers and hence shares many of the properties of the Twitter social network. Different sessions of a broadcaster, therefore, tend to follow similar statistics due to the effects of social selection and peer influence~\cite{LGK12}. It was shown in~\cite{HGK14} that live sessions on live online gaming platforms can be viewed as \textit{communities} and communities in online social media have similar information consumption patterns~\cite{DBAFFAGH16}. We therefore model the interest dynamics as a time homogeneous Markov chain having a transition matrix $P$. 

{\em 2.~Engagement Dynamics:} The interest in the video, $X_t$, cannot be measured directly by the broadcaster and has to be inferred from the viewer engagement, denoted by $Y_t$. 
Since the viewer engagement measures the number of likes in a given time interval, we model it using a Markov modulated Poisson distribution. 
Denote the rate of the Poisson observation process when the interest is in state $i$ by $g_i$. 
The observation probability in~\eqref{eq:obprob} can be obtained using $B(i,j) = \frac{g_i^j \exp{(-g_i)}}{j!}$. 
{\em 3.~Broadcaster Revenue:} 
The ad revenue in online social media depends on the click rate (the probability that the ad will be clicked). 
In a recent research, Adobe Research\footnote{\url{https://gigaom.com/2012/04/16/adobe-ad-research/}} concluded that video viewers are more likely to engage with an ad if they are interested in the content of the video that the ad is inserted into. 
The reward vector in Section~\ref{subsec:ad:scheduling:problem} should capture the positive correlation that exists between interest in the videos and the click rate~\cite{lehmann2012models}.
Since the information regarding the click rate and actual number of viewers are not available in the dataset, we choose the reward vector $r$ to be a vector of decreasing elements, each being proportional to the reward in that state, such that~\ref{ass:technical} is satisfied.

{\em 4.~Broadcaster operation:}
The broadcaster wishes to schedule at most $L$ ads at instants when the interest is high. Here, we choose\footnote{Most of the popular Periscope sessions last $15-30$ mins. Broadcast television usually average $13.5$ mins per hour of advertisement or approximately one ad every $5$ mins. Hence, we choose the number of advertisements $L=5$. } the number of stops $L = 5$. At each discrete time, after receiving the observation $Y_t$, the broadcaster either stops and schedules an ad or continues with the live stream; see Figure~\ref{fig:blkdiagram:live:scheduling}.  
The ad scheduling model that we consider in this paper assumes that the interest in the content does not change with scheduling ads. This is a simplified model when the live video content is paused to allow for advertisements, as in Twitch. However, the model captures the \emph{in-video overlay} ads that are popular in YouTube Live. In video overlay ads, the advertisement is shown in a portion of the screen (typically below). Here, it is safe to assume that the interest is not affected by ad-scheduling. 

{\em 5.~Broadcaster objective: }
The objective of the broadcaster is given by~\eqref{eq:discountedreward_h}. 
It aims to schedule ads when the content is interesting, so as to elicit maximum number of clicks, thereby maximizing the expected revenue.
In personalized live streaming applications like Periscope, the discount factor in~\eqref{eq:discountedreward_h} captures the ``impatience'' of live broadcaster in scheduling ads. 

The above model and formulation correspond to a multiple stopping problem with $L$ stops, as discussed in Section~\ref{sec:system:model}. 
In the next section, we describe how to estimate the model parameters from the data (viewer engagement $Y_t$) for computing the linear threshold policies using Algorithm~\ref{algo:policy:gradient:algorithm} in Section~\ref{sec:spsa:MLR:threshold}. 

\begin{figure}
	  \includegraphics[scale=0.45]{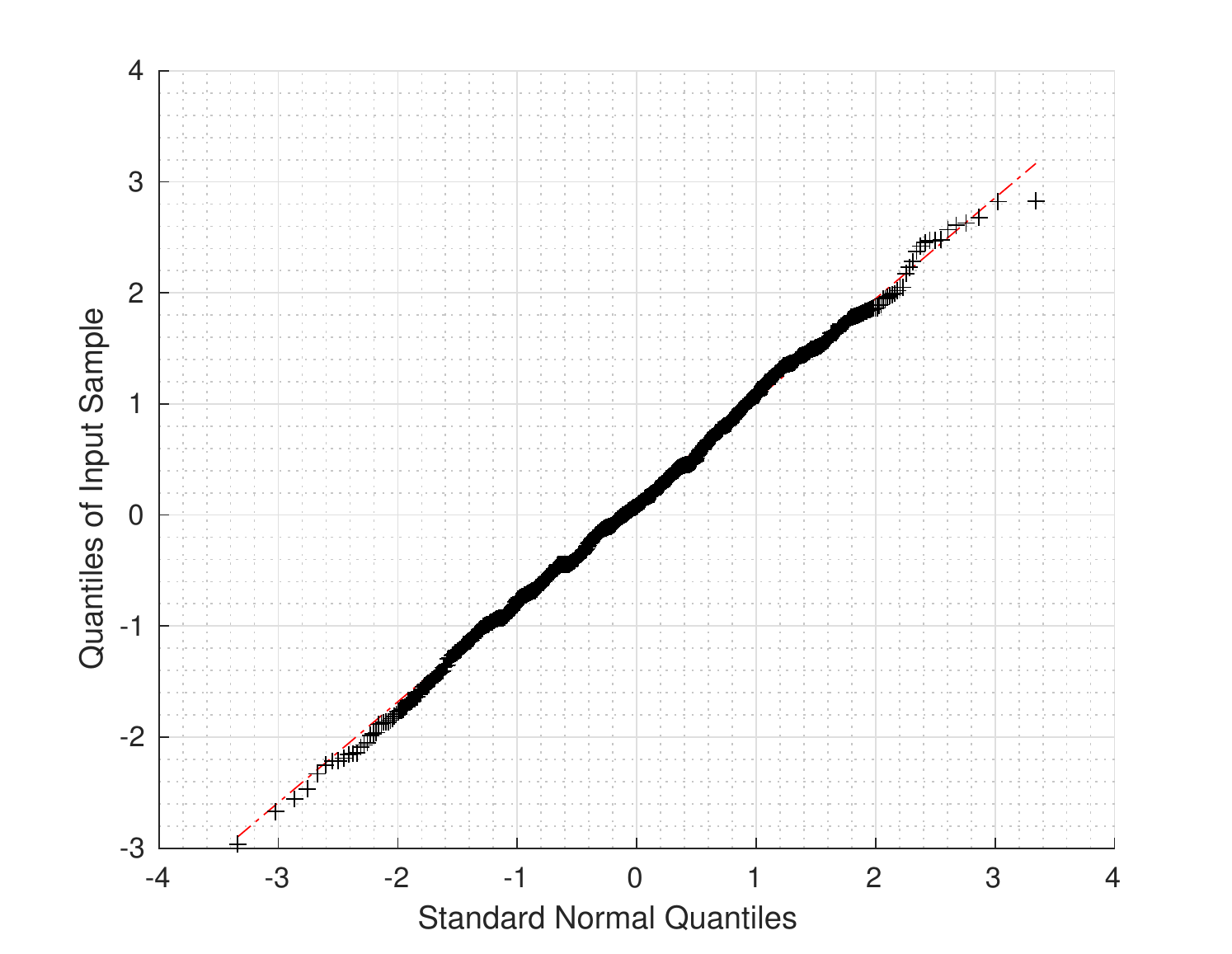}
	  \label{fig:qqplot:youtube} 
  	\caption[Periscope]{The maximum likelihood estimated parameters are given in~\eqref{eqn:realdata:study}. The QQ-plot is used for validating the goodness of fit. The linearity of the points suggest that the estimated parameters in~\eqref{eqn:realdata:study} are a good fit. }
	\label{fig:youtube:live} 
\end{figure}

{\bf Estimation of parameters: }
The live video sessions in Periscope have a range of $10-20$~minutes~\cite{WZWZZ16}. 
The viewer engagement information consists of a time series of \textit{likes} obtained by sampling the timestamped likes at a $2$-second interval. 
Sampling at a $2$-second interval, each session provides $1000$ data points. 
The model parameters $P$ and $B$ are computed using maximum likelihood estimation. Since the interest dynamics are time homogeneous, we utilize data from multiple sessions to estimate the parameters $P$ and $B$. 
The model was validated using the QQ-plot (see Figure~\ref{fig:youtube:live}) of normal pseudo-residuals~\cite[Section~{6.1}]{zucchini2009hidden}.
The estimated value of the transition matrix $P$ and the state dependent mean $g$ of a popular live session are given as:  
\begin{equation}
	\begin{aligned}
	P & = 	
	\begin{bmatrix}
		0.733 & 0.266 & 0.000 & 0.000 \\
		0.081 & 0.718 & 0.201 & 0.000 \\
		0.000 & 0.214 & 0.670 & 0.116 \\
		0.000 & 0.000 & 0.222 & 0.778
	\end{bmatrix}, \\
		g &= 	\begin{bmatrix}
		38 & 21 & 10 & 1
	\end{bmatrix}.	\end{aligned}
		\label{eqn:realdata:study}
\end{equation}
The model order dimension was estimated using the penalized likelihood criterion; specifically Table~\ref{tab:bic:selection:live:session} shows the model order selection using the Bayesian information criterion (BIC). 
The likelihood values in Table~\ref{tab:bic:selection:live:session} were obtained using an Expectation-Maximization (EM) algorithm~\cite{zucchini2009hidden}. 
In Table~\ref{tab:bic:selection:live:session} that $S=4$ has the lowest BIC value. 
\begin{table}[!h]
	\centering
	\caption{BIC model order selection for the popular live session. The maximum likelihood estimated parameters are given in~\eqref{eqn:realdata:study}. The BIC criteria was run for $S$ varying from $2-12$ (only values for $2-6$ are shown below). It can be seen that $S=4$ has the lowest BIC value. \label{tab:bic:selection:live:session}}
	\begin{tabular}{|c|c|c|}
		\toprule
		$S$ & $-\log(\mathscr{L})$ & $\text{BIC} =  -2\log(\mathscr{L})+n\log(N)$\\
		\midrule
		2 &   -4707.254 & 9535.053 \\
		3 &   -4190.652 & 8601.122 \\
		\textbf{4} &   -3969.955 & \textbf{8287.364} \\
		5 &   -3951.155 & 8405.764 \\
		6 &   -3887.453 & 8462.725\\
		\bottomrule
	\end{tabular}
	\vspace{1ex}
	{\footnotesize\begin{compactitem}
	\item $\mathscr{L}$ denotes the likelihood value. 
	\item $n$ denotes the number of parameters: $n = S^2+S-1$. 
	\item $N$ denotes the number of observations. Here, $N = 10^4$.  
	\end{compactitem}
	}		
\end{table}

The reward vector was chosen as $ r = \begin{bmatrix} 4 & 3 & 2 & 1 \end{bmatrix} $, and satisfies \ref{ass:technical} for $\rho \in \left[0,1\right]$. 

\subsubsection{Multiple ad scheduling: Performance results}
We now compare the linear threshold scheduling policies (obtained from Algorithm~\ref{algo:policy:gradient:algorithm}) with two existing schemes: 
\begin{compactenum}
\item {\em Periodic: } 
	Here, the broadcaster stops periodically to advertise. Twitch\footnote{Twitch is a video platform that focuses primarily on video gaming. In 2015, Twitch had more than $1.5$~million broadcasters and $100$~million visitors per month}, for example, uses periodic ad scheduling~\cite{SOW13}. 
Periodic advertisement scheduling is also widely used for pre-recorded videos on social media platforms like YouTube. 
\item {\em Heuristic: } Here, the broadcaster solves a classical single stopping problem at each stop. 
	The scheduler re-initializes and solves for $L=1$ in Section~\ref{sec:system:model} at each stop. \end{compactenum}
{\bf Performance Results: }
It was seen that the optimal linear threshold policies outperforms conventional periodic scheduling by $25\%$ and the heuristic scheduling by $10\%$. The periodic scheme performs poorly because it does not take into account the viewer engagement or the interest in the content while scheduling ads.  
The multiple stopping policy, in comparison to the heuristic scheme, takes into account the fact that $L$-ads need to be scheduled and hence, is optimal. 

\subsection{Large state space models \& Comparison with SARSOP}
To illustrate the application on large state space models, we present a numerical example using synthetic data.  

{\em POMDP parameters:} We consider a Markov chain with $25$ states. 
The transition matrix and observation distribution are generated as discussed in~\cite{KR14}. 
In order for the transition matrix $P$ satisfy the TP2 assumption in~\ref{ass:transition}, we use the following approach: First construct a $5$-state transition matrix $A = \exp(Qt)$, where $Q$ is a tridiagonal generator matrix (off-diagonal entries are non-negative and row sums to $0$) and $t > 0$. Since Kronecker product preserves TP2 structure, we let $P = A \otimes A$. The observation distribution $B$, containing $25$ observations satisfying~\ref{ass:observation} is similarly generated. The reward vector is chosen as follows: $r=\left[25,24,\cdots,1\right]$. The number of stops is $L= 5$. 

Because of the large state space dimension, computing the optimal policy using dynamic programming is intractable. We compare linear threshold policies (obtained through Algorithm~\ref{algo:policy:gradient:algorithm}), the heuristic policy and periodic policy (described in the Section~\ref{subsec:real:data:youtube:twitch}), in terms of the expected cumulative reward by each of the policy. Also, we compare the linear threshold policy against the state-of-the-art solver for POMDP: SARSOP (an approximate POMDP planning algorithm)~\cite{KHLSARSOP}. 

Table~\ref{tab:comparision:algorithm} shows the normalized cumulative reward by each of the policies. The expected reward was calculated using $1000$ independent Monte Carlo simulations. 
From Table~\ref{tab:comparision:algorithm} we observe the following:
\begin{compactenum}
\item The linear threshold policy and heuristic policy outperforms periodic scheduling by a factor of $2$. 
\item The linear threshold policy outperforms the heuristic policy by $12\%$. 
\item The linear threshold policy has a performance drop of $9\%$ compared to the solution obtained using SARSOP. This can be attributed to the linear hyperplane approximation to the threshold curve compared to the SARSOP solution where the number of linear segments is exponential in the number of states and observations.
\end{compactenum}

Although the linear threshold policies have a slight performance drop compared to SARSOP, it has two significant advantages:
\begin{compactenum}
\item The policy (the linear threshold vectors corresponding to each stop) is easy to implement\footnote{The SARSOP policy has approximately $\expnumber{4}{4}$ linear segments. }. 
\item Computing the linear threshold approximation is computationally cheaper compared to SARSOP algorithm. It can be noted from Table~\ref{tab:comparision:algorithm} that Algorithm~\ref{algo:policy:gradient:algorithm} is computationally cheaper by a factor of $10$. 
\end{compactenum}
\begin{table}[!h]
	\centering
	\begin{tabular}{c|c|c}
																					\toprule
		Algorithm &  Cumulative Reward & \#Computations\\
				\midrule
		SARSOP &  1 & $\expnumber{18}{12}$\\
		Linear Threshold  & 0.91 & $\expnumber{1.25}{11}$\\
		Heuristic & 0.79 & $\expnumber{1.25}{11}$\\
		Periodic &  0.35 & 0\\
		\bottomrule
	\end{tabular}
	\caption{Comparison of the expected cumulative reward (Normalized w.r.t SARSOP) and number of computations by various algorithm. The linear threshold policies have a performance drop of $9\%$ compared to the solution obtained using SARSOP and outperforms the heuristic policy by $12\%$. \\	
		SARSOP solution computed using a $2.5$~{GHz} CPU running for $2$ hours. The calculation assumes a floating point operation every CPU cycle. Algorithm~\ref{algo:policy:gradient:algorithm}, for obtaining linear threshold policies, was run with finite horizon $N=1000$.  \label{tab:comparision:algorithm}}
\end{table}

For the multiple stopping problem, Table~\ref{tab:comparision:algorithm} shows that linear threshold policies that exploit the structure of the optimal policy perform nearly as well as the optimal policy computed via a general purpose approximate POMDP solver, with a magnitude lower computational cost.
\section{Conclusion}
\label{sec:conclusion}
We presented four main results regarding the multiple stopping time problem. \\
\begin{inparaenum}[(i)]
\item The optimal policy was shown to be monotone with respect to a specialized monotone likelihood ratio order on lines (under reasonable conditions). Therefore the optimal policy was characterized by multiple threshold curves on the belief space and the optimal stopping sets satisfied a nested property (Theorem~\ref{thm:main}). \\
\item The cumulative reward was shown to be monotone with respect to the copositive ordering of the transition matrix (Theorem~\ref{thm:copositive:reward}). \\
\item Necessary and sufficient conditions were given for linear threshold policies to satisfy the MLR increasing condition for the optimal policy (Theorem~\ref{thm:coefficent:MLR:condition} and Theorem~\ref{thm:constraints:parameter}). We then gave a stochastic gradient algorithm (Algorithm~\ref{algo:policy:gradient:algorithm}) to estimate the linear threshold policies. \\
\item Finally, the linear scheduling policy was illustrated on a real data set involving interactive advertising in live social media videos. 
\end{inparaenum}

Extension of the current work could involve developing upper and lower myopic bounds to the optimal policy as in~\cite{KP15}, optimizing the ad length, and constraints on ad placement in the advertisement scheduling problem. 
\appendix
\renewcommand{\thesection}{Appendix \Alph{section}}
\renewcommand{\thesubsection}{\Alph{section}.\arabic{subsection}}
\renewcommand\theequation{\Alph{section}.\arabic{equation}}

\section{Preliminaries and Definitions}
\label{appendix:def:mlr:lines}
Theorem~\ref{thm:main} require concepts in stochastic dominance~\cite{KR80} and submodularity~\cite{topkis2011supermodularity}. 
\subsection{First-order and MLR stochastic dominance}
\label{subsec:mlr:ordering}
In order to compare belief states, we will use the monotone likelihood ratio (MLR) stochastic ordering and a specialized version of the MLR order restricted to lines in the simplex. The MLR stochastic order is useful since it is preserved under conditional expectations. 
\begin{definition}[MLR ordering]
  	\label{def:mlr:ordering}
	Let $\pi_1 ,\pi_2 \in \Pi$ be two belief state vectors.  
	Then, $\pi_1$ is greater than $\pi_2$ with respect to Monotone Likelihood Ratio (MLR) ordering--denoted as $\pi_1 \ge_{r} \pi_2$, if
	\begin{equation}
		\pi_1(j) \pi_2(i) \le \pi_2(j)\pi_1(i), \; i < j, \; i, j \in \left\{1,\dots,S\right\} 		\label{eqn:mlr:ordering}
	\end{equation}
\end{definition}
\begin{definition}[First order stochastic dominance]
	Let $\pi_1, \pi_2 \in \Pi$ be two belief state vectors. 
	Then, $\pi_1$ is greater than $\pi_2$ with respect to first-order stochastic dominance--denoted as $\pi_1 \ge_{s} \pi_2$, if
	\begin{equation}
		\sum_{i = j}^S \pi_1(i) \le \sum_{i=j}^S \pi_2(i) \quad \forall j \in \left\{1,2,\cdots,S\right\}. 
		\label{eqn:stochastic:dominance:ordering}
	\end{equation}
\end{definition}
{\em Result~\cite{krishnamurthy2016partially}:}
\begin{compactenum}[i)]
	\item $\pi_1, \pi_2 \in \Pi$. Then, $\pi_1 \ge_r \pi_2$ implies $\pi_1 \ge_{s} \pi_2$. 
	\item $\pi_1 \ge_s \pi_2$ if and only if for any increasing function $\phi(\cdot)$, $\E_{\pi_1}\left\{\phi(x)\right\} \ge \E_{\pi_2}\left\{\phi(x)\right\}$. 	
\end{compactenum}
For state-space dimension $S=2$, MLR is a complete order and coincides with first-order stochastic dominance. For state-space dimension $S > 2$ MLR is a \emph{partial order} i.e.\ $[\Belief, \ge_{r}]$ is a partially ordered set\footnotemark since it is not always possible to order any two belief states. However, on line segments in the simplex defined below, MLR is a total ordering.
\footnotetext{A \textbf{partially ordered set} is a set $X$ on which there is a binary relation $\preccurlyeq$ that is reflexive, antisymmetric, and transitive. }

Define the sub simplex $\mathcal{H}_i; i=1,S$ as: \begin{equation}
	\mathcal{H}_i = \left\{\bar{\pi}:\bar{\pi} \in \Pi \text{ and } \bar{\pi}(i) = 0\right\}.
	\label{eqn:def:H}
\end{equation}
Figure~\ref{fig:mlr:lines} illustrates $\mathcal{H}_1$ for $S=3$. 
Consider two types of lines, $\mathcal{L}\left(e_i,\bar{\pi}\right); i = 1,S$, where $e_i$ is the unit indicator vector with $1$ in the $i$ position and $0$ elsewhere, as follows:
For any $\bar{\pi} \in \mathcal{H}_i$, construct the line $\mathcal{L}(e_i,\bar{\pi})$ that connects $\bar{\pi}$ to $e_i$ as below:
	\begin{equation}
					\mathcal{L}\left(e_i,\bar{\pi}\right) =  \left\{\pi \in \Pi: \pi = \left(1-\gamma\right)\bar{\pi} + \gamma e_1,  0\le \gamma \le 1\right\}, 
			{\bar{\pi} \in \mathcal{H}_1}
			\label{eqn:def:L:e1}
			\end{equation}
				With an abuse of notation, we denote $\mathcal{L}(e_i,\bar{\pi})$ by $\mathcal{L}(e_i)$.
Figure~\ref{fig:mlr:lines} illustrates the definition of $\mathcal{L}(e_1)$. 
\begin{definition}[MLR ordering on lines]
	$\pi_1$ is greater than $\pi_2$ with respect to MLR ordering on the lines $\mathcal{L}(e_i)$, denoted as $\pi_1 \ge_{\mathcal{L}_i} \pi_2$, if $\pi_1, \pi_2 \in \mathcal{L}(e_i)$, for some $\bar{\pi} \in \mathcal{H}_i$ and $\pi_1 \ge_{r} \pi_2$. 
\end{definition}
\begin{remark}[\cite{krishnamurthy2016partially}]\normalfont
	For $i = 1,S$, $\pi_1 \ge_{\mathcal{L}_i} \pi_2$ is equivalent to $\pi_j = \varepsilon_j e_i + (1-\varepsilon_j) \bar{\pi}$, for some $\bar{\pi} \in \mathcal{H}_i$ and $\varepsilon_1 \ge \varepsilon_2$. 
	\label{remark:explicit:chara}
\end{remark}
{\underline{\em Discussion:}} The MLR ordering on lines is a complete order, i.e.\ it forms a \emph{chain}, i.e.\ all elements $\pi_1, \pi_2 \in \mathcal{L}(e_i)$ are comparable, i.e.\ either $\pi_1 \ge_{\mathcal{L}_i} \pi_2$ or $\pi_2 \ge_{\mathcal{L}_i} {\pi_1}$. To see why this is the case, if $\pi_1 \ge_{\mathcal{L}_1} \pi_2$, then from the definition of MLR dominance, the element wise ratio $\frac{\pi_1}{\pi_2}(i)$ should be decreasing in $i$. It is easy to see that
\begin{equation}
	\cfrac{\pi_1}{\pi_2}(i)= \begin{cases} \frac{\varepsilon_1}{\varepsilon_2} & i=1 \\ \frac{1-\varepsilon_1}{1-\varepsilon_2} & i\ne1 \end{cases}.
\end{equation}
Hence, if $\varepsilon_1 \ge \varepsilon_2$ then $\pi_1 \ge_{\mathcal{L}_i} \pi_2$. Hence, the MLR ordering of probability vectors $\pi_1$ and $\pi_2$ reduces to the scalar ordering of $\varepsilon_1$ and $\varepsilon_2$, which is fully ordered. Similar argument holds when $\pi_1, \pi_2 \in \mathcal{L}(e_{S},\bar{\pi})$. The intuition for why it only works for $\mathcal{L}(e_{1},\bar{\pi})$ and $\mathcal{L}(e_{S},\bar{\pi})$ is that the trick only works when $\frac{\varepsilon_1}{\varepsilon_2}$ is at either end, i.e. either when $i=1$ or $i=S$. The complete order on $\mathcal{L}(e_{i},\bar{\pi}); i =1,S$ allows us to give a threshold characterization of the optimal policy on the belief space. \\
\begin{definition}[TP2]
	A stochastic matrix, $A$ is Totally Positive of order 2 (TP2), if all the second order minors are non-negative i.e.\ the determinants
	\begin{equation}
		\begin{vmatrix}
			a_{i_1j_1} & a_{i_1j_2} \\
			a_{i_2j_1} & a_{i_2j_2}
		\end{vmatrix}
		\ge 0, \forall i_2\ge i_1, j_2\ge j_1
		\label{eqn:definition:TP2}
	\end{equation}
	\label{def:TP2:ordering}
\end{definition}
Equivalently, for any row index $i$, $A_{i,j}/A_{i+1,j}$ is increasing in $j$. For a continuous distribution, let $A_{(i)}$ denote the probability density while the Markov chain is in state $i$. Then $A$ being TP2 is equivalent to $\cfrac{A_{(j)}(x)}{A_{(i)}(x)}$ being a non-decreasing function of $x$. 

An important consequence of assumption~\ref{ass:transition} and~\ref{ass:observation} is the following theorem, which state that the filter $\filter(\belief,y)$ in~\eqref{eq:hmmfilter} preserves MLR dominance. 
\begin{theorem}[\cite{krishnamurthy2016partially}]
	If the transition matrix, $P$, and the observation matrix, $B$, satisfies the condition in~\ref{ass:transition} and~\ref{ass:observation}, then 
	\begin{compactitem}
	\item For $\pi_1 \ge_r \pi_2$, the filter satisfies $\filter(\belief_1,\cdot) \ge_r \filter(\belief_2,\cdot)$. 
	\item For $\pi_1 \ge_r \pi_2$, $\sigma(\pi_1,\cdot) \ge_s \sigma(\pi_2,\cdot)$  
	\end{compactitem}
	\label{thm:VK:filter}
\end{theorem}
To prove the structural result, we show that the $Q(\pi,l,u)$ in~\eqref{eq:bellman} is submodular on the lines $\mathcal{L}(e_i); i=1,S$ with respect to the MLR order $\ge_{\mathcal{L}_{i}}$. 
\begin{definition}[Submodular function]
	A function $f: \mathcal{L}(e_i) \times \left\{1,2\right\} \rightarrow \reals$ is submodular if : 
	\begin{equation}
		\hspace{-.51cm}f(\pi,u) - f(\pi,\bar{u}) \le f(\bar{\pi},u) - f(\bar{\pi},\bar{u}); u \ge \bar{u}, \pi \ge_{\mathcal{L}_{i}} \bar{\pi}
		\label{eqn:def:submodular}
	\end{equation}
	\label{def:submodular}
\end{definition}
\begin{theorem}[\cite{topkis2011supermodularity}]
				If $f(\pi,u)$ is submodular, then there exists a $u^*(\pi) =  \underset{u \in \mathcal{U}}{\argmax} f(\pi,u)$ that is decreasing in $\pi$. 
		\end{theorem}
\section{Value Iteration}
\label{subsec:value:iteration}
The value iteration algorithm is a successive approximation approach for solving Bellman's equation~\eqref{eq:bellman}. However, in this paper, we use the value iteration algorithm in a mathematical induction proof; and not as a numerical algorithm. For iterations $k = 0,1,\dots$, \begin{equation}
	\label{eqn:dyn}
	V_{k+1}(\pi,l) = \underset{u \in \{1,2\}}{\text{max}}  Q_{k+1}(\pi,l,u), 
\end{equation}
\begin{equation}
	\label{eqn:dyn:policy}
	\mu_{k+1}(\pi,l) = \underset{u \in \{1,2\}}{\argmax }  Q_{k+1}(\pi,l,u), 
\end{equation}
where
\begin{equation}
	Q_{k+1}(\pi,l,1) = \reward{\pi} + \rho \sum_{y} V_{k}(\Tpiy,l-1)\Spiy, 
	\label{eqn:Q:stop} \end{equation}
\begin{equation}
	Q_{k+1}(\pi,l,2) = \rho \sum_{y} V_{k}(\Tpiy,l)\Spiy,  
	\label{eqn:Q:continue}
\end{equation}
with $V_{0}(\pi,l)$ initialized arbitrarily. 
Define $W_{k}(\pi,l)$ as 
\begin{equation}
	W_{k}(\pi,l) \triangleq V_{k}(\pi,l)-V_{k}(\pi,l-1).
	\label{eqn:def:W:k}
\end{equation}
The stopping and continue sets (at each iteration $k$) when $l$ stops are remaining is defined as follows:
\begin{align}
	\begin{aligned}
		{S_{k+1}^{l}} = \{ \pi | \reward{\pi} \geq \rho \sum_y W_{k}(\Tpiy,l)\Spiy \},  \\
		{C_{k+1}^{l}} = \{ \pi | \reward{\pi} < \rho \sum_y W_{k}(\Tpiy,l)\Spiy \}.
	\end{aligned}
\label{eqn:continueset}
\end{align} 
The optimal stationary policy $\mu^*(\pi,l)$ is given by 
\begin{equation}
	\mu^*(\pi,l) = \lim_{k \to \infty} \mu_{k}(\pi,l).
	\label{eqn:value:iteration:limit:policy}
\end{equation}
Correspondingly, the stationary stopping and continue sets in~\eqref{eqn:defn:stopping:region} and~\eqref{eqn:defn:continue:region} are given by
\begin{equation}
	S^l = \lim_{k \to \infty} S_{k}^l, \quad C^l = \lim_{k \to \infty} C_{k}^l.
	\label{eqm:value:iteration:limit:sets}
\end{equation}
The value function, $V_{k}(\pi,l)$ in~\eqref{eqn:dyn}, can be rewritten, using~\eqref{eqn:continueset}, as follows:
	\begin{align}
		V_{k}(\pi,l) 	& = {\left(\reward{\pi} + \rho \sum_{y} V_{k-1}(\Tpiy,l-1)\Spiy\right)\mathcal{I}_{S_{k}^{l}}} \nonumber\\
				& + {\left(\rho \sum_{y} V_{k-1}(\Tpiy,l)\Spiy\right)\mathcal{I}_{C_{k}^{l}}}, 
	\label{eqn:V:k:pi:l}
	\end{align}
where $\mathcal{I}_{C_{k}^{l}}$ and $\mathcal{I}_{S_{k}^{l}}$ are indicator functions on the continue and stopping sets respectively, for each iteration $k$. 

Assume $S_{k}^{l-1} \subset S_{k}^{l}$ (see Theorem~\ref{prop:nested}) and substituting~\eqref{eqn:V:k:pi:l} in the definition of $W_{k}(\pi,l)$ in~\eqref{eqn:def:W:k}, 
\begin{align}
  W_{k}(\pi,l)  &= \left(\rho \sum_{y} W_{k-1}(\Tpiy,l)\Spiy\right) \mathcal{I}_{C_{k}^{l}}(\pi) \nonumber \\
  		&+ {\reward{\pi} \mathcal{I}_{C_{k}^{l-1}\cap S_{k}^{l}}(\pi)} \label{w_k_l} \\
		&+ \left(\rho \sum_{y} W_{k-1}(\Tpiy,l-1)\Spiy\right) \mathcal{I}_{S_{k}^{l-1}}(\pi). \nonumber
\end{align}

In order to prove the main theorem (Theorem~\ref{thm:main}), we require the following results, proofs of which are provided in~\ref{appendix:proofs}. 
\begin{theorem}
	\label{prop:V:increase:pi}
	$V_{k}(\pi,l)$ is increasing in $\pi$.
\end{theorem}

\begin{theorem}
	\label{prop:W:decrease:l}
	$W_{k}(\pi,l)$ is decreasing in $l$.
\end{theorem}

\begin{theorem}
	\label{prop:nested}
	$S_{k+1}^l \supset S_{k+1}^{l-1}$
\end{theorem}
\section{Proof of Theorems~1 and~2}
\label{appendix:proofs}
We prove Theorem~\ref{prop:V:increase:pi}, Theorem~\ref{prop:W:decrease:l} and Theorem~\ref{prop:nested} using induction and assume that the theorems hold for all values less than $k$. \subsection{Proof of Theorem~\ref{prop:V:increase:pi}}
Recall from~\eqref{eqn:dyn}, 
\begin{equation*}
	V_{k}(\pi,l) = \underset{u \in \{1,2\}}{\text{max}}  Q_{k}(\pi,l,u), 
\end{equation*}
To prove Theorem~\ref{prop:V:increase:pi}, we show $Q_{k}(\pi,l,u)$ is MLR increasing in $\pi$ for $u=\left\{1,2\right\}$. 

Recall from~\eqref{eqn:Q:stop}, 
\begin{equation*}
	\scalebox{0.94}[1]{$Q_{k}(\pi,l,1) = \reward{\pi} + \rho \sum_{y} V_{k-1}(\Tpiy,l-1)\Spiy$}, 
\end{equation*}
Using Theorem~\ref{thm:VK:filter} and the induction hypothesis, the term $\sum_{y} V_{k-1}(\Tpiy,l-1)\Spiy$ is MLR increasing in $\pi$. From Assumption~\ref{ass:technical}, $\reward{\pi}$ is MLR increasing in $\pi$. The proof for $Q_{k}(\pi,l,2)$ MLR increasing in $\pi$ is similar and is omitted. Hence, $V_{k}(\pi,l)$ is MLR increasing in $\pi$. 
\subsection{Proof of Theorem~\ref{prop:W:decrease:l}}
  The proof follows by induction. 
  Recall from~\eqref{w_k_l}, we have
	  		  \begin{align}
		  		  W_{k}(\pi,l-1) & = 	\sum_{y} {W_{k-1}(\Tpiy,l-1)\Spiy \mathcal{I}_{C_{k}^{l-1}}(\pi)} + \nonumber\\
	  		 	& {\reward{\pi} \mathcal{I}_{C_{k}^{l-2}\cap S_{k}^{l-1}}(\pi)} +  \label{w_k_lm1} \\
				& {\sum_{y} W_{k-1}(\Tpiy,l-2)\Spiy \mathcal{I}_{S_{k}^{l-2}}(\pi)} \nonumber
		\end{align}
  	  Hence, we compare $W_{k}(\pi,l)$ and $W_{k}(\pi,l-1)$ in the following $4$ regions:
	\begin{compactitem}
	\item[a.)] $S_{k}^{l-2}: W_{k}(\pi,l) - W_{k}(\pi,l-1) = $
	    \begin{equation*}
		    \sum_{y} ( W_{k-1}(\Tpiy,l-1) - W_{k-1}(\Tpiy,l-2) )\Spiy,
	    \end{equation*}
	    which is non-negative by the induction assumption. 
    \item[b.)] 	 $C_{k}^{l-2} \cap S_{k}^{l-1}: W_{k}(\pi,l) - W_{k}(\pi,l-1) = $  
	    \begin{equation*}
		    \sum_{y} W_{k-1}(\Tpiy,l-1) \Spiy  - \reward{\pi},
	    \end{equation*}
	    which is non-negative since $\scalebox{0.9}[1]{$\pi \in S_k^{l-1}$}$. 
	  \item[c.)] $C_{k}^{l-1} \cap S_{k}^{l}: W_{k}(\pi,l) - W_{k}(\pi,l-1) = $ 
	    \begin{equation*}
		    \reward{\pi} - \sum_{y} W_{k-1}(\Tpiy,l-1) \Spiy,
	    \end{equation*}
	    which is non-negative since $\scalebox{0.9}[1]{$\pi \in C_k^{l-1}$}$. 
	  \item[d.)] $C_{k}^{l}: W_{k}(\pi,l) - W_{k}(\pi,l-1) = $ 
	    \begin{equation*}
		    \sum_{y} ( W_{k-1}(\Tpiy,l) - W_{k-1}(\Tpiy,l-1) )\Spiy,
	    \end{equation*}
	    which is non-negative by the induction assumption.  	\end{compactitem}  
\subsection{Proof of Theorem~\ref{prop:nested}}
If $\pi \in S_{k}^{l-1}$, then $\reward{\pi} \ge \sum_{y} W_{k-1}(\Tpiy,l-1)\Spiy$. By Theorem~\ref{prop:W:decrease:l}, $\reward{\pi} \ge \sum_{y} W_{k-1}(\Tpiy,l)\Spiy$. Hence $\pi \in S_{k}^l$. 
\subsection{Proof of Theorem~\ref{thm:main}}
\label{subsec:proof:thm:main}
\textbf{Existence of optimal policy:} 
In order to show the existence of a threshold policy of $\mathcal{L}(e_1)$, we need to show that $Q_{k+1}(\pi,l,2) - Q_{k+1}(\pi,l,1)$ is submodular in $\pi \in \mathcal{L}(e_1)$. 
Since, 
\begin{equation*}
	Q_{k+1}(\pi,l,2) - Q_{k+1}(\pi,l,1) =  \rho \sum_{y} W_{k}(\Tpiy,l)\Spiy  - \reward{\pi}.
	\end{equation*}
We need to show that $\rho \sum_{y} W_{k}(\Tpiy,l)\Spiy  - \reward{\pi}$ is MLR decreasing in $\pi$. 
\begin{align}
	& \rho \sum_{y} W_{k}(\Tpiy,l)\Spiy  - \reward{\pi}  \\
	& = \sum_{y} \left( \rho W_{k}(\Tpiy,l)  - \reward{\pi} \right) \Spiy \nonumber \\
	& = \sum_{y} \left( \left(\rho W_{k}(\Tpiy,l)-\rho \reward{\Tpiy}\right) \right. \nonumber \\
	& \left.- \left(\reward{\pi} -\rho \reward{\Tpiy}\right) \right) \Spiy \nonumber \\
	& = \rho \sum_{y} \left(W_{k}(\Tpiy,l)-\reward{\Tpiy}\right) \Spiy \nonumber \\ 
	& - r^\prime (I - \rho P^\prime) \pi  \label{eqn:sufficent:W:decreasing}
\end{align}
The term $-r^\prime (I - \rho P^\prime) \pi$ in~\eqref{eqn:sufficent:W:decreasing} is MLR decreasing in $\pi$ due to our assumption. 
Hence, to show that $\rho \sum_{y} W_{k}(\Tpiy,l)\Spiy  - \reward{\pi}$ is MLR decreasing in $\pi$ it is sufficient to show that $W_{k}(\pi,l)  - \reward{\pi}$ is MLR decreasing in $\pi$. 
Define, 
\begin{equation}
	\bar{W}_{k}(\pi,l) \triangleq  W_{k}(\pi,l)  - \reward{\pi}
	\label{eqn:define:W:bar}
\end{equation}

Now, $\bar{W}_{k}(\pi,l) = $  
\begin{align}
	& \scalebox{0.90}[1]{$\left(\sum_{y} \rho \left( \left(\bar{W}_{k-1}(\Tpiy,l) + \reward{\Tpiy} \right) - \reward{\pi}\right) \Spiy \right) \mathcal{I}_{C_{k}^{l}}(\pi)  + $}\nonumber \\
	& 
	\scalebox{0.90}[1]{$\left(\sum_{y} \rho \left( \left(\bar{W}_{k-1}(\Tpiy,l-1) + \reward{\Tpiy} \right) - \reward{\pi}\right) \Spiy \right) \mathcal{I}_{S_{k}^{l}}(\pi) $}\nonumber \\
	&
	\scalebox{0.90}[1]{$= \left(\sum_{y} \left(\rho \bar{W}_{k-1}(\Tpiy,l) \Spiy \right) -r^\prime (I -\rho P)^\prime \pi \right)  \mathcal{I}_{C_{k}^{l}}(\pi)  + $}\nonumber \\
	& 
	\scalebox{0.90}[1]{$\left(\sum_{y} \left(\rho \bar{W}_{k-1}(\Tpiy,l-1)\Spiy \right) -r^\prime (I -\rho P)^\prime \pi \right)  \mathcal{I}_{S_{k}^{l}}(\pi) $}\label{eqn:W:bar:inductive:step}
\end{align}
We prove using induction that $\bar{W}_{k}(\pi,l)$ is MLR decreasing in $\pi$, using the recursive relation over $k$ in~\eqref{eqn:W:bar:inductive:step}. 

For $k=0$,
\begin{equation}
	\bar{W}_{0}(\pi,l) = W_{0}(\pi,l)-r^\prime \pi 
	 = V_{0}(\pi,l)-V_{0}(\pi,l-1)-r^\prime \pi 
	 \label{eqn:W:0:pi:l}
	\end{equation}
The initial conditions of the value iteration algorithm can be chosen such that $\bar{W}_{0}(\pi,l)$ in~\eqref{eqn:W:0:pi:l} is decreasing in $\pi$. A suitable choice of the initial conditions is given below: 
\begin{equation}
	V_{0}(\pi,l) = r^\prime \left(\sum_{j=0}^{l-1} \rho^j P^j\right)^\prime \pi.
	\label{eqn:value:iteration:initial:conditions}
\end{equation}
The intuition behind the initial conditions in~\eqref{eqn:value:iteration:initial:conditions} is that the value function, $V_{0}(\pi,l)$ gives the expected total reward if we stop $l$ times successively starting at belief $\pi$. 

Next, we show that $\bar{W}_{k}(\pi,l)$ is MLR decreasing in $\pi$, if $\bar{W}_{k-1}(\pi,l)$ is MLR decreasing in $\pi$. 
For $\pi_1 \ge_r \pi_2$, consider the following cases: 
\begin{inparaenum}[(a)] 
\item \label{thm:casea} $\pi_1, \pi_2 \in S_{k}^{l-1}$, 
\item \label{thm:caseb} $\pi_1 \in S_{k}^{l-1}$, $\pi_2 \in C_{k}^{l-1}\cap S_{k}^{l}$, 
\item \label{thm:casec} $\pi_1, \pi_2 \in C_{k}^{l-1}\cap S_{k}^{l}$, 
\item \label{thm:cased} $\pi_1 \in C_{k}^{l-1}\cap S_{k}^{l}$, $\pi_2 \in C_{k}^{l}$, 
\item \label{thm:casee} $\pi_1,\pi_2 \in C_{k}^{l}$, 
\item \label{thm:casef} $\pi_1 \in S_{k}^{l-1}$, $\pi_2 \in C_{k}^{l}$. 
\end{inparaenum}
For cases~\eqref{thm:casea}, \eqref{thm:casec}, \eqref{thm:casee}, $\bar{W}_{k}(\pi_1,l) \le \bar{W}_{k}(\pi_2,l)$ by the induction assumption. For case~\eqref{thm:caseb} $\bar{W}_{k}(\pi_1,l) \le \bar{W}_{k}(\pi_2,l)$, since $\pi_1 \in S_{k}^{l-1}$. Case~\eqref{thm:cased} is similar to case~\eqref{thm:caseb}. For case~\eqref{thm:casef}, 
\begin{align*}
&\bar{W}_{k}(\pi_1,l) - \bar{W}_{k}(\pi_2,l) 	\\ 
&= \scalebox{0.95}{$\left(\sum_{y} \left(\rho \bar{W}_{k-1}(T(\pi_1,y),l-1)\sigma(\pi_1,y) \right) -r^\prime (I -\rho P)^\prime \pi_1 \right)$} \\
&- \scalebox{0.95}{$\left(\sum_{y} \left(\rho \bar{W}_{k-1}(T(\pi_2,y),l) \sigma(\pi_2,y) \right) -r^\prime (I -\rho P)^\prime \pi_2 \right)$} \\
&\scalebox{0.95}{$\le \rho \left(\sum_{y} \left( \left(\bar{W}_{k-1}(T(\pi_1,y),l-1) - \bar{W}_{k-1}(T(\pi_1,y),l) \right)\sigma(\pi_1,y) \right)\right)$} \\
&\le 0,
\end{align*}
where the first inequality is due to induction hypothesis and the second inequality is due to Theorem~\ref{prop:W:decrease:l}.
Hence, it is clear that $\bar{W}_{k}(\pi,l)$ is decreasing in $\pi$, if $\bar{W}_{k-1}(\pi,l)$ is decreasing in $\pi$, finishing the induction step. 

\textbf{Characterization of the switching curve $\Gamma_l$:} For each $\bar{\pi} \in \mathcal{H}$ construct the line segment $\mathcal{L}(e_1,\bar{\pi})$. 
The line segment can be described as $(1-\varepsilon) \bar{\pi} + \varepsilon e_1$. On the line segment $\mathcal{L}(e_1,\bar{\pi})$ all the belief states are MLR orderable. 
Since $\mu^*(\pi,l)$ is monotone decreasing in $\pi$, for each $l$, we pick the largest $\varepsilon$ such that $\mu^*(\pi,l)=1$. The belief state, $\pi^{\varepsilon^*, \bar{\pi}}$ is the threshold belief state, where $\varepsilon^* = \text{inf }\{\varepsilon \in [0,1]: \mu^*({\pi^{\varepsilon,\bar{\pi}}}) = 1\}$. Denote by $\Gamma(\bar{\pi}) = \pi^{\varepsilon^*, \bar{\pi}}$. 
The above construction implies that there is a unique threshold $\Gamma(\bar{\pi})$ on $\mathcal{L}(e_1,\bar{\pi})$. The entire simplex can be covered by considering all pairs of lines $\mathcal{L}(e_1,\bar{\pi})$, for $\bar{\pi} \in \mathcal{H}_1$, i.e. $\Pi = \cup_{\bar{\pi} \in \mathcal{H}} \mathcal{L}(e_1,\bar{\pi})$. Combining, all points yield a unique threshold curve in $\Pi$ given by $\Gamma = \cup_{\bar{\pi} \in \mathcal{H}_1} \Gamma(\bar{\pi})$. 

\textbf{Connectedness of $S^l$:} Since $e_1 \in S^l$ for all $l$, call $S_a^l$, the subset of $S^l$ that contains $e_1$. 
Suppose $S_b^l$ is the subset that was disconnected from $S_a^l$. 
Since every point on $\Pi$ lies on the line segment $\mathcal{L}(e_1,\bar{\pi})$, for some $\bar{\pi}$, there exists a line segment starting from $e_1 \in S^l_a$ that would leave the set $S_a^l$, pass through the set where action $2$ is optimal and then intersect set $S_b^l$, where action $1$ is optimal. 
But, this violates the requirement that the policy $\mu^*(\pi,l)$ is monotone on $\mathcal{L}(e_1,\bar{\pi})$. 
Hence, $S_a^l$ and $S_b^l$ are connected. 

\textbf{Connectedness of $C^l$:} Assume $e_S \in C^l$, otherwise $C^l$ is empty and there is nothing to prove. 
Call the set that contains $e_S$ as $C^l_a$. 
Suppose $C_b^l \subset C^l$ is disconnected from $C_a^l$. 
Since every point in $\Pi$ lies on the line segment $\mathcal{L}(e_S,\bar{\pi})$, for some $\bar{\pi}$, there exists a line starting from $e_S \in C^l_a$ would leave set $C_a^l$, pass through the set where action $1$ is optimal and then intersect the set $C_b^l$ (where action $2$ is optimal). But this violates the monotone property of $\mu^*(\pi,l)$.  

\textbf{Nested structure:} The proof is straightforward from Theorem~\ref{prop:nested}. 
\subsection{Copositive ordering and Proof of Theorem~\ref{thm:copositive:reward}}
\begin{definition}[Copositive ordering]
	\label{def:copositive:ordering:transition:matrices}
	Given two $S \times S$ transition matrices $P$ and $Q$, we say that $$P \preceq Q$$ if the sequence of $S \times S$ matrices $\Gamma^{j}; j = 1,\cdots,S-1$ are:
	$$\pi^\prime \Gamma^{j} \pi \ge 0, \quad \pi \in \Pi, \quad \text{for each $j$,}$$
	where each element of $\Gamma^{j}$ is given by:
	$$\Gamma^{j}_{m,n} = \frac{1}{2}\left(\gamma_{m,n}^{j} + \gamma_{n,m}^{j}\right), $$ and $$\gamma_{m,n}^{j} = P_{m,j} Q_{n,j+1} - P_{m,j+1} Q_{n,j}.$$
\end{definition}

A consequence of copositive ordering is the following theorem
\begin{theorem}[{\cite[Theorem~{10.6.1}]{krishnamurthy2016partially}}]
	\label{thm:10:6:1}
	Suppose transition matrices $\underbar{$P$}$ and $\bar{P}$ are constructed such that $\underbar{$P$} \preceq P \preceq \bar{P}$. Then for any observation $y$ and belief $\pi \in \Pi$, the filtering update $T(\pi,y;P)$\footnote{The notation $T(\cdot,\cdot;P)$ makes explicit the transition matrix used in the filter update.} in~\eqref{eq:hmmfilter} satisfies 
	\begin{equation*}
		T(\pi,y;\underbar{$P$}) \le_{r} T(\pi,y;P) \le_{r} T(\pi,y;\bar{P}).
	\end{equation*}
\end{theorem}

\subsubsection{Proof of Theorem~\ref{thm:copositive:reward}}
\label{proof:copositive:theorem}
We prove that dominance of the transition matrix (in terms of copositive ordering) $P \succeq  \bar{P}$ results in dominance of the rewards, i.e.\ $V(\pi,l;P) \ge V(\pi,l;\bar{P})$. The proof follows by induction. 
For $n=0$, $V_n(\pi,l;P) \ge V_n(\pi,l;\bar{P})$ by suitable initialization of the value iteration algorithm. 
Next, to prove the inductive step assume $V_n(\pi,l;P) \ge V_n(\pi,l;\bar{P})$. 
By the induction hypothesis and Theorem~\ref{thm:10:6:1}, 
\begin{equation}
	V_n(T(\pi,y;\bar{P}),l;P) \ge V_n(T(\pi,y;\bar{P}),l;\bar{P})
	\label{eqn:V:bound:1}
\end{equation}
By Theorem~\ref{prop:V:increase:pi} under assumptions~\ref{ass:transition} to~\ref{ass:technical}, $V_n(\pi;P)$ and $V_n(\pi;\bar{P})$ are MLR increasing in $\pi$. 
Hence, 
\begin{equation}
	V_n(T(\pi,y;P),l;P) \ge V_n(T(\pi,y;\bar{P}),l;P)
	\label{eqn:V:bound:2}
\end{equation}
Combining~\eqref{eqn:V:bound:1} and~\eqref{eqn:V:bound:2} gives $V_n(T(\pi,y;P),l;P) \ge V_n(T(\pi,y;\bar{P}),l;\bar{P})$. 

Under Assumption~\ref{ass:transition} to~\ref{ass:observation} and Theorem~\ref{thm:VK:filter} we have $T(\pi,y;P)$ is MLR increasing in $y$. Hence, $V_n(T(\pi,y;P),l;P)$ is MLR increasing in $y$. Under Assumption~\ref{ass:transition} to~\ref{ass:observation}, since $T(\pi,y;P) \ge_{r} T(\pi,y;\bar{P})$ implies $\sigma(\pi,y;P) \ge_{s} \sigma(\pi,y;\bar{P})$ we have,
\begin{equation}
	\begin{aligned}
	& \reward{\pi} + \rho \sum_{y} V_{n}(T(\pi,y;P),l-1) \sigma(\pi,y;P) \ge \\
	&\reward{\pi} + \rho \sum_{y} V_{n}(T(\pi,y;\bar{P}),l-1) \sigma(\pi,y;\bar{P})
	\end{aligned}
	\label{eqn:co:1}
\end{equation}
and
\begin{equation}
	\begin{aligned}
	&\rho \sum_{y} V_{n}(T(\pi,y;P),l) \sigma(\pi,y;P) \ge \\
	&\rho \sum_{y} V_{n}(T(\pi,y;\bar{P}),l) \sigma(\pi,y;\bar{P})
	\end{aligned}
	\label{eqn:co:2}
\end{equation}
Maximizing both sides of equation~\eqref{eqn:co:1} and~\eqref{eqn:co:2} gives $V_{n+1}(T(\pi,y;P),l;P) \ge V_{n+1}(T(\pi,y;\bar{P}),l;\bar{P})$, finishing the induction step. Theorem~\ref{thm:copositive:reward} follows by substituting $l=L$ in $V(\pi,l;P) \ge V(\pi,l;\bar{P})$.  
\subsection{Proof of Theorem~\ref{thm:coefficent:MLR:condition}}
 \label{appendix:proof:thm:lines:MLR:condition}
 The proof of Theorem~\ref{thm:coefficent:MLR:condition} is similar to the proof of Theorem~{12.4.1} in~\cite{krishnamurthy2016partially}. Recall, that the linear threshold policies is given by: 
 \begin{equation*}
 	\mu_\theta(\pi, l) = \begin{cases}
		1 & \text{if } \begin{bmatrix} 0 & 1 & \theta_l \end{bmatrix} \begin{bmatrix} \pi \\ -1\end{bmatrix} \le 0 \\
		2 & \text{else }.
	\end{cases}
	\label{eqn:threshold:policy:proof}
\end{equation*}
For any number of stops remaining, $e_1$ (the belief that the state is $1$) belongs to the stopping set, $S^l$
,which gives the first condition $\theta_l(S-1) \ge 0$. 

Consider $\pi_1 \ge_{\mathcal{L}_1} \pi_2$. Then $\pi_1 = \varepsilon_1 e_1 + (1-\varepsilon_1) \bar{\pi}$ and $\pi_2 = \varepsilon_2 e_1 + (1-\varepsilon_2) \bar{\pi}$, for some $\bar{\pi} \in \mathcal{H}$ and $\varepsilon_1 \ge \varepsilon_2$\footnote{Refer to Remark~\ref{remark:explicit:chara}}. 
For the linear policy to the MLR decreasing on lines, $\mu_\theta(\pi_1, l) \le \mu_\theta(\pi_2, l)$. 
Hence,
\begin{equation*}
		\begin{bmatrix} 0 & 1 & \theta_l \end{bmatrix} \begin{bmatrix} \pi_1 \\ -1\end{bmatrix}  \le \begin{bmatrix} 0 & 1 & \theta_l \end{bmatrix} \begin{bmatrix} \pi_2 \\ -1\end{bmatrix} ,
\end{equation*}
\begin{equation*}
		\begin{bmatrix} 0 & 1 & \theta_l \end{bmatrix} \begin{bmatrix} \pi_1 - \pi_2 \\ 0\end{bmatrix}  \le 0 ,
\end{equation*}
\begin{equation*}
		\begin{bmatrix} 0 & 1 & \theta_l \end{bmatrix} \begin{bmatrix} (\varepsilon_1 - \varepsilon_2) e_1  - (\varepsilon_1 - \varepsilon_2) \bar{\pi}  \\ 0 \end{bmatrix}  \le 0 ,
\end{equation*}
\begin{equation*}
	-(\varepsilon_1 - \varepsilon_2)\left[\bar{\pi}(2) + \theta_l(1) \bar{\pi}(3) + \cdots + \theta_l(S-2) \bar{\pi}(S)\right] \le 0,
\end{equation*}
giving the second set of conditions $\theta_l(i) \ge 0,\; i \le S-2$. 

The proof of the second part is similar and hence is omitted. 
\subsection{Proof of Theorem~\ref{thm:constraints:parameter}}
\label{appendix:proof:thm:constraints:parameter}
For $l_1 > l_2$, due to the nested structure in Theorem~\ref{thm:main} $S^{l_2} \subset S^{l_1}$. 
This implies the following
\begin{align}
	\mu_\theta(l_2,\pi) &\ge \mu_\theta(l_1,\pi) \nonumber \\
	\begin{bmatrix} 0& 1& \theta_{l_2} \end{bmatrix} \begin{bmatrix} \pi \\ -1 \end{bmatrix} &\ge \begin{bmatrix} 0& 1& \theta_{l_1} \end{bmatrix} \begin{bmatrix} \pi \\ -1 \end{bmatrix} \nonumber \\
	\begin{bmatrix} 0& 0& \theta_{l_2}-\theta_{l_1} \end{bmatrix} \begin{bmatrix} \pi \\ -1 \end{bmatrix} &\ge 0 \label{eqn:final:step:proof:subsetting:structure}
\end{align}
It is straightforward to check that the conditions in~\eqref{eqn:cond:parameter} in Theorem~\ref{thm:constraints:parameter} satisfy the conditions in~\eqref{eqn:final:step:proof:subsetting:structure}. 
\section{Proof of Propositions}
\subsection{Proof of Proposition~\ref{prop:r:decreasing:elements}}
\label{proof:prop:r:decreasing:elements}
The proof follows in two steps. Let $v = (I-\rho P)r$. 

When $\rho < 1$, $(I-\rho P)$ is invertible. Hence, $r = (I-\rho P)^{-1} v = \sum_{k=0}^\infty \rho^k P^k v$. Since the product of TP2 matrices is TP2, each $P^k$ is TP2. Then, $r$ been decreasing follows from Theorem~{9.2.2} in~\cite{krishnamurthy2016partially}. 

For $\rho = 1$, $g = \lim_{\rho \uparrow 1} (1- \rho)(I - \rho P)^{-1}  v$ is the solution of $(I-P)r = v$. This limit exists~\cite[Cor.~{8.2.5}]{puterman} and hence, $r$ has decreasing elements. 
\subsection{Proof of Proposition~\ref{prop:stopping:set:convex:union}}
\label{proof:prop:stopping:set:convex:union}
The proof follows from the finite stopping time property of the multiple stopping time problem; see Footnote~\ref{foot:undiscount}. A finite horizon POMDP with a finite state and observation space has a value function that is piecewise linear and convex; see Theorem~{7.4.1} in \cite{krishnamurthy2016partially}. For $l=1$, 
\begin{equation*}
V(\pi,1) = \underset{\gamma \in \Gamma}{\max} \gamma^\prime \pi,
\end{equation*}
where $\Gamma$ is a finite set due to the finite stopping time property. 
For $l=2$, the dynamic programming equation in~\eqref{eq:bellman} can be written as: 
\begin{equation*}
V(\pi,2) = \max\left\{ r^\prime \pi +  \underset{\gamma \in \Gamma}{\max} \gamma^\prime P^\prime \pi, \discount\sum_{y \in \obspace}\valuefunction\left(\filter(\belief,y),2\right)\filternorm(\belief,y)\right\}. 
\end{equation*}
For each $\gamma \in \Gamma$, the stopping set is convex; see the proof of Theorem~{12.2.1} in~\cite{krishnamurthy2016partially}. Hence, the stopping set for $l=2$ is a union of convex sets. Similar argument holds for any value of $l$. 
\section{Finite Horizon Approximation Algorithms}
\label{appendix:finite:time:approximation:algorithm}
Algorithm~\ref{algo:JN} details the steps to compute the finite time horizon approximation in~\eqref{eqn:finite:time:approximation} for the linear threshold policies. Algorithm~\ref{algo:JN} takes as input the POMDP parameters, policy (in terms of the parameter $\theta$) and the number of stops. It computes the accumulated reward using the input policy by running a POMDP simulation of at most $N$ time points. 
\begin{algorithm}
\caption{Finite Horizon Approximation Algorithm for linear threshold policies}
\label{algo:JN}
\begin{algorithmic}[1]
              \Require Finite time approximation parameter $N$, policy parameter $\theta$, number of stops $L$, initial belief $\pi_0$, discount factor $\rho$, reward vector $r$.  
       \State $l \gets L$, $J \gets 0$. 
       \For{iterations $n=1,2,\cdots N$: }
       \While{$l \ne 0$}
       \State Obtain observation $Y_n$ and update belief $\pi_n$ according to~\eqref{eq:hmmfilter}. 
       \State Compute $a_n \gets \mu_\theta(\pi_n,l)$ according to~\eqref{eqn:threshold:policy}. 
       \If{$a_n = 1$}
       \State $J \gets J + \rho^n \pi_n^\prime r$ 
       \State $l \gets l -1$
       \EndIf
       \EndWhile
       \EndFor
       \Return $J$
\end{algorithmic}
\end{algorithm}

Algorithm~\ref{algo:JN:softmax} summarizes the computation of the finite time horizon approximation with the softmax parametrization of the policy in~\eqref{eqn:soft-max}. The key difference of Algorithm~\ref{algo:JN:softmax} with Algorithm~\ref{algo:JN} is in Steps~{5-7}. In Steps~{5-7} of Algorithm~\ref{algo:JN:softmax} the softmax policy in~\eqref{eqn:soft-max} replaces the linear threshold policies in Step~5 of Algorithm~\ref{algo:JN}. 
\begin{algorithm}
\caption{Finite Horizon Approximation Algorithm: Using softmax parametrization}
\label{algo:JN:softmax}
\begin{algorithmic}[1]
	\Require Finite time approximation parameter $N$, policy parameter $\theta_{l,u}$, number of stops $L$, initial belief $\pi_0$, discount factor $\rho$, reward vector $r$.  
       \State $l \gets L$, $J \gets 0$. 
       \For{iterations $n=1,2,\cdots N$: }
       \While{$l \ne 0$}
       \State Obtain observation $Y_n$ and update belief $\pi_n$ according to~\eqref{eq:hmmfilter}. 
              \State $\scalebox{0.95}{$\text{actionprob} = \left[\exp{\left(\begin{bmatrix} 0 & \theta_{l,1}\end{bmatrix}^\prime \pi\right)} \; \exp{\left(\begin{bmatrix} 0 & \theta_{l,2}\end{bmatrix}^\prime \pi\right)}\right]$}$
       \State $\text{actionprob} \gets \text{actionprob}/\sum\text{actionprob}$
       \State Sample $a_n \sim \text{actionprob}$
       \If{$a_n = 1$}
       \State $J \gets J + \rho^n \pi_n^\prime r$ 
       \State $l \gets l -1$
       \EndIf
       \EndWhile
       \EndFor
       \Return $J$
\end{algorithmic}
\end{algorithm}

\begin{thebibliography}{10}

\bibitem{Lai97}
T.~L. Lai, ``On optimal stopping problems in sequential hypothesis testing,''
  {\em Statistica Sinica}, vol.~7, no.~1, pp.~33--51, 1997.

\bibitem{Lai01}
T.~L. Lai, {\em Sequential analysis}.
\newblock Wiley Online Library, 2001.

\bibitem{Rus87}
J.~Rust, ``Optimal replacement of gmc bus engines: An empirical model of harold
  zurcher,'' {\em Econometrica: Journal of the Econometric Society},
  pp.~999--1033, 1987.

\bibitem{Mon80}
G.~E. Monahan, ``Optimal stopping in a partially observable {M}arkov process
  with costly information,'' {\em Operations Research}, vol.~28, no.~6,
  pp.~1319--1334, 1980.

\bibitem{poor2009quickest}
H.~V. Poor and O.~Hadjiliadis, {\em Quickest Detection}.
\newblock Cambridge University Press, 2008.

\bibitem{Kri11}
V.~Krishnamurthy, ``Bayesian sequential detection with phase-distributed change
  time and nonlinear penalty -- {A} {POMDP} {L}attice programming approach,''
  {\em IEEE Transactions on Information Theory}, vol.~57, pp.~7096--7124, Oct
  2011.

\bibitem{KB16}
V.~Krishnamurthy and S.~Bhatt, ``{S}equential {D}etection of {M}arket {S}hocks
  with {R}isk-{A}verse {CVaR} {S}ocial {S}ensors,'' {\em IEEE Journal of
  Selected Topics in Signal Processing}, vol.~10, pp.~1061--1072, Sept 2016.

\bibitem{krishnamurthy2016partially}
V.~Krishnamurthy, {\em Partially Observed {M}arkov Decision Processes}.
\newblock Cambridge University Press, 2016.

\bibitem{nak85}
T.~Nakai, ``The problem of optimal stopping in a partially observable {M}arkov
  chain,'' {\em Journal of Optimization Theory and Applications}, vol.~45,
  no.~3, pp.~425--442, 1985.

\bibitem{BBM04}
S.~Bollapragada, M.~R. Bussieck, and S.~Mallik, ``Scheduling commercial
  videotapes in broadcast television,'' {\em Oper. Res.}, vol.~52,
  pp.~679--689, Oct. 2004.

\bibitem{PC15}
D.~G. Popescu and P.~Crama, ``Ad revenue optimization in live broadcasting,''
  {\em Management Science}, vol.~62, no.~4, pp.~1145--1164, 2015.

\bibitem{KM11}
H.~Kang and M.~P. McAllister, ``Selling you and your clicks: examining the
  audience commodification of google,'' {\em Journal for a Global Sustainable
  Information Society}, vol.~9, no.~2, pp.~141--153, 2011.

\bibitem{Kle05}
R.~Kleinberg, ``A multiple-choice secretary algorithm with applications to
  online auctions,'' in {\em Proceedings of the sixteenth annual ACM-SIAM
  symposium on Discrete algorithms}, pp.~630--631, Society for Industrial and
  Applied Mathematics, 2005.

\bibitem{sta87}
W.~Stadje, ``An optimal k-stopping problem for the poisson process,'' in {\em
  Mathematical Statistics and Probability Theory}, pp.~231--244, Springer,
  1987.

\bibitem{Nik99}
M.~Nikolaev, ``On optimal multiple stopping of {M}arkov sequences,'' {\em
  Theory of Probability \& Its Applications}, vol.~43, no.~2, pp.~298--306,
  1999.

\bibitem{Ann15}
A.~Krasnosielska-Kobos, ``Multiple-stopping problems with random horizon,''
  {\em Optimization}, vol.~64, no.~7, pp.~1625--1645, 2015.

\bibitem{ER15}
E.~Bayraktar and R.~Kravitz, ``Quickest detection with discretely controlled
  observations,'' {\em Sequential Analysis}, vol.~34, no.~1, pp.~77--133, 2015.

\bibitem{GBL14}
J.~Geng, E.~Bayraktar, and L.~Lai, ``Bayesian quickest change-point detection
  with sampling right constraints,'' {\em IEEE Transactions on Information
  Theory}, vol.~60, no.~10, pp.~6474--6490, 2014.

\bibitem{NJ10}
S.~H.~J. Alexander G.~Nikolaev, ``Stochastic sequential decision-making with a
  random number of jobs,'' {\em Operations Research}, vol.~58, no.~4,
  pp.~1023--1027, 2010.

\bibitem{ST05}
S.~Savin and C.~Terwiesch, ``Optimal product launch times in a duopoly:
  Balancing life-cycle revenues with product cost,'' {\em Operations Research},
  vol.~53, no.~1, pp.~26--47, 2005.

\bibitem{LPVZ15}
I.~Lobel, J.~Patel, G.~Vulcano, and J.~Zhang, ``Optimizing product launches in
  the presence of strategic consumers,'' {\em Management Science}, vol.~62,
  no.~6, pp.~1778--1799, 2015.

\bibitem{WRA11}
K.~E. Wilson, R.~Szechtman, and M.~P. Atkinson, ``A sequential perspective on
  searching for static targets,'' {\em European Journal of Operational
  Research}, vol.~215, no.~1, pp.~218 -- 226, 2011.

\bibitem{CT08}
R.~Carmona and N.~Touzi, ``Optimal multiple stopping and valuation of swing
  options,'' {\em Mathematical Finance}, vol.~18, no.~2, pp.~239--268, 2008.

\bibitem{DL15}
E.~Dahlgren and T.~Leung, ``An optimal multiple stopping approach to
  infrastructure investment decisions,'' {\em Journal of Economic Dynamics and
  Control}, vol.~53, pp.~251--267, 2015.

\bibitem{bertsekas1995dynamic}
D.~P. Bertsekas, {\em Dynamic programming and optimal control}, vol.~1.
\newblock Athena Scientific Belmont, MA, 2017.

\bibitem{PT87}
C.~H. Papadimitriou and J.~N. Tsitsiklis, ``The complexity of {M}arkov decision
  processes,'' {\em Math. Oper. Res.}, vol.~12, pp.~441--450, Aug 1987.

\bibitem{mullercomparison}
A.~M{\"u}ller and D.~Stoyan, ``Comparison methods for stochastic models and
  risks. 2002,'' {\em John Wiley\&Sons Ltd., Chichester}.

\bibitem{YZ06}
G.~Yin and Q.~Zhang, {\em Discrete-time Markov chains: two-time-scale methods
  and applications}, vol.~55.
\newblock Springer Science \& Business Media, 2006.

\bibitem{PB16}
G.~Piao and J.~G. Breslin, ``Exploring dynamics and semantics of user interests
  for user modeling on twitter for link recommendations,'' in {\em Proceedings
  of the 12th International Conference on Semantic Systems}, pp.~81--88, ACM,
  2016.

\bibitem{KR14}
V.~Krishnamurthy and C.~R. Rojas, ``Reduced complexity hmm filtering with
  stochastic dominance bounds: A convex optimization approach,'' {\em IEEE
  Transactions on Signal Processing}, vol.~62, pp.~6309--6322, Dec 2014.

\bibitem{pflug2012optimization}
G.~C. Pflug, {\em Optimization of stochastic models: the interface between
  simulation and optimization}, vol.~373.
\newblock Springer Science \& Business Media, 2012.

\bibitem{spall2005introduction}
J.~C. Spall, {\em Introduction to stochastic search and optimization:
  estimation, simulation, and control}, vol.~65.
\newblock John Wiley \& Sons, 2005.

\bibitem{LOV91}
W.~S. Lovejoy, ``A survey of algorithmic methods for partially observed
  {M}arkov decision processes,'' {\em Annals of Operations Research}, vol.~28,
  no.~1, pp.~47--65, 1991.

\bibitem{bishop2006pattern}
C.~M. Bishop, {\em Pattern recognition and machine learning}.
\newblock springer, 2006.

\bibitem{sutton1998reinforcement}
R.~S. Sutton and A.~G. Barto, {\em Reinforcement learning: An introduction},
  vol.~1.
\newblock MIT press Cambridge, 1998.

\bibitem{WZWZZ16}
B.~Wang, X.~Zhang, G.~Wang, H.~Zheng, and B.~Y. Zhao, ``Anatomy of a
  personalized livestreaming system,'' in {\em Proceedings of the 2016 Internet
  Measurement Conference}, IMC '16, (New York, NY, USA), pp.~485--498, ACM,
  2016.

\bibitem{SOW13}
T.~Smith, M.~Obrist, and P.~Wright, ``Live-streaming changes the (video)
  game,'' in {\em Proc.\ of the 11th European Conference on Interactive TV and
  Video}, pp.~131--138, ACM, 2013.

\bibitem{baldominos2016real}
A.~Baldominos~G{\'o}mez, E.~Albacete~Garc{\'\i}a, I.~Marrero, and
  Y.~Saez~Achaerandio, ``Real-time prediction of gamers behavior using variable
  order {M}arkov and big data technology: a case of study,'' 2016.

\bibitem{AMM12}
N.~Archak, V.~Mirrokni, and S.~Muthukrishnan, ``Budget optimization for online
  campaigns with positive carryover effects,'' in {\em Proc. of the 8th
  International Conference on Internet and Network Economics}, pp.~86--99,
  Springer-Verlag, 2012.

\bibitem{BRTMA09}
F.~Benevenuto, T.~Rodrigues, M.~Cha, and V.~Almeida, ``Characterizing user
  behavior in online social networks,'' in {\em Proceedings of the 9th ACM
  SIGCOMM Conference on Internet Measurement}, IMC '09, pp.~49--62, 2009.

\bibitem{LGK12}
K.~Lewis, M.~Gonzalez, and J.~Kaufman, ``Social selection and peer influence in
  an online social network,'' {\em Proceedings of the National Academy of
  Sciences}, vol.~109, no.~1, pp.~68--72, 2012.

\bibitem{HGK14}
W.~A. Hamilton, O.~Garretson, and A.~Kerne, ``Streaming on twitch: Fostering
  participatory communities of play within live mixed media,'' in {\em
  Proceedings of the 32Nd Annual ACM Conference on Human Factors in Computing
  Systems}, pp.~1315--1324, 2014.

\bibitem{DBAFFAGH16}
M.~Del~Vicario, A.~Bessi, F.~Zollo, F.~Petroni, A.~Scala, G.~Caldarelli, H.~E.
  Stanley, and W.~Quattrociocchi, ``The spreading of misinformation online,''
  {\em Proceedings of the National Academy of Sciences}, vol.~113, no.~3,
  pp.~554--559, 2016.

\bibitem{lehmann2012models}
J.~Lehmann, M.~Lalmas, E.~Yom-Tov, and G.~Dupret, ``Models of user
  engagement,'' in {\em International Conference on User Modeling, Adaptation,
  and Personalization}, pp.~164--175, Springer, 2012.

\bibitem{zucchini2009hidden}
W.~Zucchini and I.~L. MacDonald, {\em Hidden {M}arkov models for time series:
  an introduction using R}.
\newblock CRC press, 2009.

\bibitem{KHLSARSOP}
H.~Kurniawati, D.~Hsu, and W.~S. Lee, ``{SARSOP}: Efficient point-based {POMDP}
  planning by approximating optimally reachable belief spaces.,'' in {\em
  Robotics: Science and Systems.}, 2008.

\bibitem{KP15}
V.~Krishnamurthy and U.~Pareek, ``Myopic bounds for optimal policy of {POMDPs}:
  An extension of {L}ovejoy's structural results,'' {\em Operations Research},
  vol.~62, no.~2, pp.~428--434, 2015.

\bibitem{KR80}
S.~Karlin and Y.~Rinott, ``Classes of orderings of measures and related
  correlation inequalities. i. multivariate totally positive distributions,''
  {\em Journal of Multivariate Analysis}, vol.~10, no.~4, pp.~467--498, 1980.

\bibitem{topkis2011supermodularity}
D.~M. Topkis, {\em Supermodularity and complementarity}.
\newblock Princeton university press, 2011.

\bibitem{puterman}
M.~L. Puterman, {\em Markov decision processes: discrete stochastic dynamic
  programming}.
\newblock John Wiley \& Sons, 2005.

\end{thebibliography}

\end{document}